\documentclass[a4paper,12pt]{article}

\usepackage[margin=1in,dvips,a4paper,nohead]{geometry}
\usepackage[margin=1em, font=small, 
justification=centerlast, tableposition=top]{caption}

\usepackage{amsfonts,amssymb,amsmath}
\usepackage[noBBpl,slantedGreek]{mathpazo}

\usepackage{pstricks}

\psset{unit=1pt, dimen=middle, linewidth=.5pt, arrowsize=3pt 2}

\newdimen \psx
\newdimen \psy

\def\psddots (#1,#2){
  \psx #1\psxunit
  \psy #2\psyunit
  \qdisk(\psx,\psy){.7pt}
  \advance \psx -.2\psxunit
  \advance \psy .2\psyunit
  \qdisk(\psx,\psy){.7pt}
  \advance \psx .4\psxunit
  \advance \psy -.4\psyunit
  \qdisk(\psx,\psy){.7pt}
}

\def\pssddots (#1,#2){
  \psx #1\psxunit
  \psy #2\psyunit
  \qdisk(\psx,\psy){.7pt}
  \advance \psx .2\psxunit
  \advance \psy .2\psyunit
  \qdisk(\psx,\psy){.7pt}
  \advance \psx -.4\psxunit
  \advance \psy -.4\psyunit
  \qdisk(\psx,\psy){.7pt}
}

\makeatletter
\def\numberbysection{\@addtoreset{equation}{section}
        \def\theequation{\thesection.\arabic{equation}}}

\numberbysection

\def \:{\mskip .5\thinmuskip}
\def\ph {{\hbox to 0pt{\phantom{$\scriptstyle -1$}\hss}}}

\def \id{\mathop{\mathrm{id}}\nolimits}

\def\bbC {\mathbb C}

\def\bbR {\mathbb R}
\def\bbZ {\mathbb Z}

\def\calF {\mathcal F}
\def\calG {\mathcal G}

\def\calL {\mathcal L}
\def\calM {\mathcal M}
\def\calN {\mathcal N}
\def\calO {\mathcal O}
\def\calP {\mathcal P}

\mathchardef\Gamma "100
\def\gothg{\mathfrak g}
\def\gothG{\mathfrak G}
\def\gothgl{\mathfrak{gl}}

\def\rme {\mathrm e}

\def\rmi {\mathrm i}
\def\rmGL {\mathrm{GL}}

\def\rmSO {\mathrm{SO}}
\def\wt {\widetilde}
\def\Mat {\mathrm{Mat}}

\title{\bf The rational dressing for abelian twisted loop Toda systems}

\author{Kh. S. Nirov\\
\small \em Institute for Nuclear Research of the Russian Academy of
Sciences\\[-.3em]
\small \em 60th October Anniversary Prospect 7a, 117312 Moscow,
Russia\\[.3em]
A. V. Razumov\\
\small \em Institute for High Energy Physics\\[-.3em]
\small \em 142281 Protvino, Moscow Region, Russia}

\date{}

\begin{document}

\maketitle

\begin{abstract}
We consider abelian twisted loop Toda equations associated with the
complex general linear groups. The Dodd--Bullough--Mikhailov
equation is a simplest particular case of the equations under 
consideration. We construct new soliton solutions of these Toda
equations using the method of rational dressing. 
\end{abstract}

\section{Introduction}
 
The comprehensive investigation of completely integrable systems has 
at least two reasons. First, such systems serve as a testing area 
for developing various methods to solve nonlinear partial differential 
equations. And second, they possess an interesting class of solutions, 
called solitons, which have properties attractive from the point of 
view of possible physical applications. 

The two-dimensional loop Toda equations provide an illustrative and 
very rich example of completely integrable nonlinear equations, see, 
for example, the monographs \cite{LezSav92,RazSav97}. 
Different methods applicable to loop Toda equations for constructing 
their soliton-like solutions were analysed in the paper
\cite{NirRaz08}. Namely, multi-soliton solutions of abelian 
{\em untwisted} loop Toda equations associated with the general 
linear groups were explicitly constructed by means of the Hirota's 
\cite{Hir04, Hol92, ConFerGomZim93, MacMcG92, AraConFerGomZim93, 
ZhuCal93} and the rational dressing \cite{ZakSha79, Mik81} methods, 
and a direct relationship between these approaches was 
established.\footnote{It is worth to note that sometimes 
it is helpful to employ a combination of such complementary 
methods, see, for example, \cite{BueFerRaz02, AssFer07}}.

In this paper we continue the investigation of abelian loop Toda equations 
associated with the complex general linear groups 
started in the paper \cite{NirRaz08}. Here we consider abelian {\em twisted}
loop Toda equations. It is interesting that the famous 
Dodd--Bullough--Mikhailov equation is a simplest particular case of
such twisted loop Toda equations. We develop the rational dressing 
method in application to these classes of nonlinear equations and 
construct for them new soliton solutions.

\section{Loop Toda equations}

In this section, mainly following the monographs
\cite{LezSav92, RazSav97} and our papers  \cite{NirRaz07a, NirRaz07b},
we recall basic notions and introduce notations to be used below. We 
start our consideration with a Lie group $\calG$ whose Lie algebra 
$\gothG$ is endowed with a $\bbZ$-gradation,
\[
\gothG = \bigoplus_{k \in \bbZ} \gothG_k,
\qquad
[\gothG_k , \gothG_l] \subset \gothG_{k+l},
\]
and denote by $L$ such a positive integer that the grading subspaces
$\gothG_k$, where $0 < |k| < L$, are trivial. We denote by $\calG_0$
the closed Lie subgroup of $\calG$ corresponding to the zero-grade Lie
subalgebra $\gothG_0$. Then, the Toda equation associated with $\calG$
is an equation for a mapping $\Xi$ of the Euclidean plane $\bbR^2$
to $\calG_0$, explicitly of the form
\begin{equation}
\partial_+ (\Xi^{-1} \partial_- \Xi) 
= [\calF_-, \Xi^{-1} \calF_+ \Xi],
\label{e:2.1}
\end{equation}
where $\calF_-$ and $\calF_+$ are some fixed mappings of $\bbR^2$
to $\gothG_{-L}$ and $\gothG_{+L}$, respectively, satisfying the relations
\begin{equation}
\partial_+ \calF_- = 0, \qquad \partial_- \calF_+ = 0.
\label{e:2.2}
\end{equation} 
Here we use the customary notation 
$\partial_- = \partial/{\partial z^-}$,
$\partial_+ = \partial/{\partial z^+}$ 
for the partial derivatives over the standard coordinates on $\bbR^2$.
Certainly, to obtain a nontrivial Toda equation we have to assume that
the subspaces $\gothG_{-L}$ and $\gothG_{+L}$ are nontrivial.

When the Lie group $\calG_0$ is abelian, the corresponding Toda equation
is said to be {\em abelian\/}, otherwise we deal with a {\em non-abelian
Toda equation\/}.

We see that a Toda equation is specified by a choice of a $\bbZ$-gradation
of the Lie algebra $\gothG$ of $\calG$ and mappings $\calF_-$,
$\calF_+$ satisfying the
conditions~(\ref{e:2.2}). Therefore, to classify the Toda equations
associated with a Lie group $\calG$  we should classify $\bbZ$-gradations
of the Lie algebra $\gothG$ of $\calG$ up to isomorphisms.

It is essential for our purposes that the Toda equation (\ref{e:2.1}) together
with the relations (\ref{e:2.2}) are equivalent to the zero-curvature condition
for a flat connection on the trivial fiber bundle $\bbR^2 \times \calG
\rightarrow \bbR^2$. Indeed, writing the zero-curvature condition as 
the equation
\begin{equation}
\partial_- \calO_+ - \partial_+ \calO_- + [\calO_-, \calO_+] = 0
\label{e:2.3}
\end{equation}
for the $\gothG$-valued components of the flat connection under 
consideration, imposing the grading conditions
\[
\calO_- = \calO_{-0} + \calO_{-L}, \qquad
\calO_+ = \calO_{+0} + \calO_{+L},
\]
and destroying the residual gauge invariance by the condition
\[
\calO_{+0} = 0, 
\]
we bring the connection components to the form
\begin{equation}
\calO_- = \Xi^{-1} \partial_- \Xi + \calF_-, 
\qquad \calO_+ = \Xi^{-1} \calF_+ \Xi,
\label{e:2.4}
\end{equation}
and then derive the equation (\ref{e:2.1}) and the relations (\ref{e:2.2})
directly from the zero-cur\-vature condition (\ref{e:2.3}), as well as vice
versa \cite{RazSav94, RazSav97, RazSav97a}.

It follows from the equality (\ref{e:2.3}) that there is a mapping $\Phi$
of $\bbR^2$ to $\calG$ such that
\begin{equation}
\Phi^{-1} \partial_- \Phi = \calO_-, \qquad \Phi^{-1} \partial_+ \Phi =
\calO_+. \label{e:r5}
\end{equation}
We say in this situation that the connection with the components $\calO_-$
and $\calO_+$ is generated by the mapping $\Phi$. 

We consider the case where $\calG$ is a twisted loop group of a complex
classical Lie group $G$ which is defined as follows. Let $a$ be an automorphism
of $G$ satisfying the relation $a^M = \id_G$ for some positive integer 
$M$.\footnote{Here $M$ is not necessarily the order of the
automorphism $a$, but can be its arbitrary multiple.} The twisted 
loop group $\calL_{a,M}(G)$ is formed by the mappings $\chi$ of 
the unit circle $S^1$ to $G$ satisfying the equality
\[
\chi(\epsilon_M p) = a(\chi(p)),
\]
where $\epsilon_M = \rme^{2 \pi \rmi/M}$ is the $M$th principal root of unity.
We think the circle $S^1$ as consisting of complex numbers of modulus one.
The group law in $\calL_{a,M}(G)$ is defined pointwise.  The Lie algebra
of $\calL_{a,M}(G)$ is the twisted loop Lie algebra $\calL_{A,M}(\gothg)$,
where $\gothg$ is the Lie algebra of $G$ and $A$ is the automorphism of
$\gothg$ corresponding to the automorphism $a$ of $G$. The Lie algebra
$\calL_{A,M}(\gothg)$ is formed by the mappings $\xi$ of $S^1$ to $\gothg$
satisfying the equality
\[
\xi(\epsilon_M p) = A(\xi(p))
\]
with the Lie algebra operation defined pointwise. Note that the relation
$A^M = \id_{\gothg}$ is satisfied automatically.

In the paper \cite{NirRaz06} we classified a wide class of the so-called
integrable $\bbZ$-gradations with finite-dimensional grading subspaces of
the twisted loop Lie algebras of the finite-dimensional complex simple Lie
algebras. Namely, we showed that any such $\bbZ$-gradation of a loop Lie
algebra $\calL_{A,M}(\gothg)$ is conjugated by an isomorphism to the standard
$\bbZ$-gradation of another loop Lie algebra $\calL_{A',M'}(\gothg)$, where
the automorphisms $A$ and $A'$ differ by an inner automorphism of $\gothg$.

Recall that for the {\em standard $\bbZ$-gradation} of the Lie algebra
$\calL_{A,M}(\gothg)$ the grading subspaces are
\[
\calL_{A,M}(\gothg)_k = \{ \xi = \lambda^k x \in \calL_{A,M}(\gothg) \mid x
\in \gothg, \ A(x) = \epsilon_M^k x \},
\]
where by $\lambda$ we denote the restriction of the standard coordinate
on $\bbC$ to $S^1$.

It is well known that twisted loop Lie algebras defined by automorphisms
which differ by an inner automorphism are isomorphic, and really
different twisted loop Lie algebras can be labeled by the elements 
of the corresponding outer automorphism group. In particular, if $A$ 
is an inner automorphism, the loop Lie algebra $\calL_{A,M}(\gothg)$ 
is isomorphic to an untwisted loop Lie algebra
$\calL(\gothg) = \calL_{\id_{\gothg}, 1}(\gothg)$. Therefore, in this 
case a Toda equation associated with $\calL_{a,M}(G)$ and specified by 
some choice of a $\bbZ$-gradation of $\calL_{A,M}(\gothg)$ is
equivalent to a Toda equation associated with 
$\calL(G) = \calL_{\id_G,1}(G)$ and specified by the corresponding
choice of a $\bbZ$-gradation of $\calL(\gothg)$. 

Thus, to describe Toda equations associated with the loop groups 
$\calL_{a,M}(G)$, where $a$ is an inner automorphism of $G$, it 
suffices to describe the Toda equations associated with the untwisted 
loop groups~$\calL(G)$. However, due to simplicity of the standard 
$\bbZ$-gradation, to study Toda equations it is more convenient, 
instead of using one untwisted loop group $\calL(G)$ and different
$\bbZ$-gradations of $\calL(\gothg)$, to use different twisted loop groups
$\calL_{a,M}(G)$ and the standard $\bbZ$-gradation of $\calL_{A,M}(\gothg)$.
Similarly, to describe the Toda equations associated with the loop groups
$\calL_{a,M}(G)$, where $a$ is an outer automorphism of $G$ satisfying
the relation $a^M = \id_G$, it suffices to use only the standard 
$\bbZ$-gradation of the loop Lie algebras
$\calL_{A,M}(\gothg)$. Having all this in mind and slightly abusing
terminology, we say that when $a$ is an outer automorphism of $G$ then 
a Toda equation associated with $\calL_{a,M}(G)$ is a {\em twisted
loop Toda equation associated with $G$\/}, and when $a$ is an inner 
automorphism of $G$ then a Toda equation associated with 
$\calL_{a,M}(G)$ is an {\em untwisted loop Toda equation associated 
with $G$\/}.\footnote{It is common to omit the word `untwisted'.}

The group $\calL_{a,M}(G)$ and its Lie algebra $\calL_{A,M}(\gothg)$ are
infinite-dimensional manifolds. Nevertheless, using the so-called exponential
law \cite{KriMic91, KriMic97}, it is possible to write the 
zero-curvature representation of the Toda equations associated 
with $\calL_{a,M}(G)$ in terms of finite-dimensional manifolds. 
The essence of this useful law can be expressed by the canonical
identification $C^\infty(\calM, C^\infty(\calN, \calP)) 
= C^\infty(\calM \times \calN, \calP)$, where $\calM$, $\calN$ 
and $\calP$ are finite-dimensional manifolds, among which $\calN$ 
is compact.

The connection components $\calO_-$ and $\calO_+$ entering the 
equality (\ref{e:2.3}) are mappings of $\bbR^2$ to the loop Lie 
algebra $\calL_{A,M}(\gothg)$. We denote the mappings of 
$\bbR^2 \times S^1$ to $\gothg$, corresponding
to $\calO_-$ and $\calO_+$ in accordance with the exponential law,
by $\omega_-$ and $\omega_+$, and call them also the connection components.
The mapping $\Phi$ generating the connection under consideration is a mapping
of $\bbR^2$ to $\calL_{a,M}(G)$. Denoting the respective mapping of 
$\bbR^2 \times S^1$ to $G$ by $\varphi$ we can write
\begin{equation}
\varphi^{-1} \partial_- \varphi = \omega_-, \qquad 
\varphi^{-1} \partial_+ \varphi = \omega_+, \label{e:2.5}
\end{equation}
which is equivalent to the expressions (\ref{e:r5}).
Having in mind that the mapping $\varphi$ uniquely determines the connection
generating mapping $\Phi$, we say that the mapping $\varphi$ also generates the
connection under consideration. We introduce, according to the exponential law,
the smooth mapping $\gamma$ of  $\bbR^2 \times S^1$ to $G$ respective to
$\Xi$, and smooth mappings of $\bbR^2 \times S^1$ to $\gothg$ respective
to $\calF_-$ and $\calF_+$. 

Now, explicitly describing the grading subspaces of the standard 
$\bbZ$-gradation of the loop Lie algebra $\calL_{A,M}(\gothg)$,
we find that the subalgebra $\calL_{A,M}(\gothg)_0$ is isomorphic
to the subalgebra $\gothg_{[0]_M}$ of $\gothg$, and the Lie group
$\calL_{a,M}(G)_0$ is isomorphic to the connected Lie subgroup $G_0$
of $G$ corresponding to the Lie algebra $\gothg_{[0]_M}$. Here 
$\gothg_{[k]_M}$ mean the grading subspaces of the $\bbZ_M$-gradation 
of $\gothg$ induced by the automorphism $A$, and $[k]_M$ denotes the 
element of the ring $\bbZ_M$ corresponding to the integer $k$. For 
the connection components $\omega_-$ and $\omega_+$ we can write the 
expressions 
\begin{equation}
\omega_- = \gamma^{-1} \partial_- \gamma + \lambda^{-L} c_-, 
\qquad
\omega_+ = \lambda^L \gamma^{-1} c_+ \gamma, \label{e:2.6}
\end{equation}
which are equivalent to the equalities (\ref{e:2.4}). Here
$c_-$ and $c_+$ are mappings of $\bbR^2$ to $\gothg_{-[L]_M}$
and $\gothg_{+[L]_M}$ respectively. Hence, the Toda equation (\ref{e:2.1}) 
can be written as
\begin{equation}
\partial_+(\gamma^{-1} \partial_- \gamma) 
= [c_-, \gamma^{-1} c_+ \gamma], \label{e:2.7}
\end{equation}
and the conditions (\ref{e:2.2}) imply that
\begin{equation}
\partial_+ c_- = 0, \qquad \partial_- c_+ = 0. \label{e:2.8}
\end{equation}
We call an equation of the form (\ref{e:2.7}) also a loop Toda equation.

Our classification of loop Toda equations is based on a convenient
block-matrix representation of the grading subspaces \cite{NirRaz07a,
NirRaz07b} we have implemented. Each element $x$ of the complex classical Lie 
algebra $\gothg$ under consideration is treated as a $p \times p$ 
block matrix $(x_{\alpha\beta})$, where $x_{\alpha \beta}$ is 
an $n_\alpha \times n_\beta$ matrix. The sum of the positive 
integers $n_\alpha$ is the size $n$ of the matrices representing 
the elements of $\gothg$. For the case of Toda equations associated with 
the loop groups $\calL_{a, M}(\rmGL_n(\bbC))$, where $a$ is an inner 
automorphism of $\rmGL_n(\bbC)$, the integers $n_\alpha$ are
arbitrary. For the other cases they should satisfy some restrictions 
dictated by the structure of the Lie algebra $\gothg$.

The mapping $\gamma$ has the block-diagonal form
\begin{equation}
\psset{xunit=1.7em, yunit=1.2em}
\gamma = \left( \raise -1.8\psyunit \hbox{\begin{pspicture}(.6,.6)
(4.5,4.2)
\rput(1,4){$\Gamma_1$}
\rput(2,3){$\Gamma_2$}
\qdisk(2.7,2.3){.7pt} \qdisk(3,2){.7pt} \qdisk(3.3,1.7){.7pt}
\rput(4,1){$\Gamma_p$}
\end{pspicture}} \right).
\label{e:2.12}
\end{equation}
For each $\alpha = 1, \ldots, p$ the mapping $\Gamma_\alpha$ is a
mapping of $\bbR^2$ to the Lie group $\rmGL_{n_\alpha}(\bbC)$. For the
case of Toda equations associated with the loop groups $\calL_{a, M}
(\rmGL_n(\bbC))$, where $a$ is an inner automorphism of
$\rmGL_n(\bbC)$, the mappings $\Gamma_\alpha$ are arbitrary. For the
other cases they satisfy some additional restrictions.

The mapping $c_+$ has the following block-matrix structure:
\begin{equation}
\psset{xunit=2.5em, yunit=1.4em}
c_+ = \left( \raise -2.4\psyunit \hbox{\begin{pspicture}(.6,.5)
(5.6,5.3)
\rput(1,5){$0$} \rput(2,4.92){$C_{+1}$}
\rput(2,4){$0$}
\qdisk(2.8,4.2){.7pt} \qdisk(3,4){.7pt} \qdisk(3.2,3.8){.7pt}
\qdisk(2.8,3.2){.7pt} \qdisk(3,3){.7pt} \qdisk(3.2,2.8){.7pt}
\qdisk(3.8,3.2){.7pt} \qdisk(4,3){.7pt} \qdisk(4.2,2.8){.7pt}
\rput(4,2){$0$} \rput(5,1.87){$C_{+(p-1)}$}
\rput(1,.94){$C_{+0}$} \rput(5,1){$0$}
\end{pspicture}} \right),
\label{e:2.13}
\end{equation}
where for each $\alpha = 1, \ldots, p-1$ the mapping $C_{+\alpha}$ is
a mapping of $\bbR^2$ to the space of $n_\alpha \times n_{\alpha+1}$
complex matrices, and $C_{+0}$ is a mapping of $\bbR^2$ to the space
of $n_p \times n_1$ complex matrices. The mapping $c_-$ has a similar
block-matrix structure:
\begin{equation}
\psset{xunit=2.5em, yunit=1.4em}
c_- = \left( \raise -2.4\psyunit \hbox{\begin{pspicture}(.5,.5)
(5.5,5.3)
\rput(1,5){$0$} \rput(5,4.92){$C_{-0}$}
\rput(1,4){$C_{-1}$} \rput(2,4){$0$}
\qdisk(1.8,3.2){.7pt} \qdisk(2,3){.7pt} \qdisk(2.2,2.8){.7pt}
\qdisk(2.8,3.2){.7pt} \qdisk(3,3){.7pt} \qdisk(3.2,2.8){.7pt}
\qdisk(2.8,2.2){.7pt} \qdisk(3,2){.7pt} \qdisk(3.2,1.8){.7pt}
\rput(4,1.87){$0$}
\rput(4,.94){$C_{-(p-1)}$} \rput(5,1){$0$}
\end{pspicture}} \right),
\label{e:2.14}
\end{equation}
where for each $\alpha = 1, \ldots, p-1$ the mapping $C_{-\alpha}$ is
a mapping of $\bbR^2$ to the space of $n_{\alpha+1} \times n_\alpha$
complex matrices, and $C_{-0}$ is a mapping of $\bbR^2$ to the space of
$n_1 \times n_p$ complex matrices. The conditions (\ref{e:2.8}) imply
\begin{equation}
\partial_+ C_{-\alpha} = 0, \qquad \partial_- C_{+\alpha} = 0, 
\qquad \alpha = 0, 1, \ldots, p-1.
\label{e:2.15}
\end{equation}
For the case of Toda equations associated with the loop groups
$\calL_{a, M}(\rmGL_n(\bbC))$, where $a$ is an inner automorphism of
$\rmGL_n(\bbC)$, the mappings $C_{\pm \alpha}$ are arbitrary. For the
other cases they should satisfy some additional restrictions.

The Toda equation (\ref{e:2.7}) is equivalent to the following system 
of equations for the mappings
$\Gamma_\alpha$:
\begin{align}
\partial_+ \left( \Gamma_1^{-1} \: \partial_- \Gamma_1^{} \right)
&= - \Gamma_1^{-1} C_{+1}^{} \: \Gamma_2^{} \: C_{-1}^{}
+ C_{-0}^{} \Gamma_p^{-1} C_{+0}^{} \Gamma_1^{},
\notag \\*
\partial_+ \left( \Gamma_2^{-1} \: \partial_- \Gamma_2^{} \right)
&= - \Gamma_2^{-1} C_{+2}^{} \: \Gamma_3^{} \: C_{-2}^{}
+ C_{-1}^{} \Gamma_1^{-1} C_{+1}^{} \Gamma_2^{},
\notag \\*
& \quad \vdots
\label{e:2.16} \\*
\partial_+ \left(\Gamma_{p-1}^{-1} \: \partial_-
\Gamma_{p-1}^{}\right)
&= - \Gamma_{p-1}^{-1} C_{+(p-1)}^{} \: \Gamma_p^{} \: C_{-(p-1)}^{}
+ C_{-(p-2)}^{} \Gamma_{p-2}^{-1} C_{+(p-2)}^{} \Gamma_{p-1}^{},
\notag \\*
\partial_+ \left( \Gamma_p^{-1} \: \partial_- \Gamma_p^{} \right)
&= - \Gamma_p^{-1} C_{+0}^{} \: \Gamma_1^{} \: C_{-0}^{}
+ C_{-(p-1)}^{} \Gamma_{p-1}^{-1} C_{+(p-1)}^{} \Gamma_p^{}.
\notag
\end{align}
It appears that in the case under consideration without any loss of
generality we can assume that the positive integer $L$, entering the
construction of Toda equations, is equal to $1$. Note also that if any
of the mappings $C_{+\alpha}$ or $C_{-\alpha}$ is a zero mapping, then
the equations (\ref{e:2.16}) are equivalent to a Toda equation associated 
with a finite-dimensional group or to a set of two such equations.

\section{Abelian Toda equations associated with loop groups of
complex general linear groups}

It can be shown that there are three types of abelian loop
Toda equations associated with the groups $\mathrm{GL}_n(\bbC)$,
see, for example the paper \cite{NirRaz08}.

\subsection{First type: untwisted loop Toda equations} \label{s:3.1}

The abelian Toda equations of the first type arise when the
automorphism $A$ is defined by the equality
\[
A(x) = h x h^{-1}, \qquad x \in \gothgl_n(\bbC),
\]
where $h$ is a diagonal matrix with the diagonal matrix elements
\begin{equation}
h_{kk} = \epsilon_n^{n-k+1}, \qquad k = 1, \ldots, n. \label{e:3.1}
\end{equation}
The corresponding automorphism $a$ of $\rmGL_n(\bbC)$ is defined by
the equality
\[
a(g) = h g h^{-1}, \qquad g \in \rmGL_n(\bbC),
\]
where again $h$ is a diagonal matrix determined by the relation
(\ref{e:3.1}). Here the integer $M$ is equal to $n$, and $A$ 
is an inner automorphism which generates a $\bbZ_n$-gradation 
of $\gothgl_n(\bbC)$. The block-matrix structure related to this
gradation is the matrix structure itself. In other words, all blocks
are of size one by one. The mappings $\Gamma_\alpha$ are mappings of
$\bbR^2$ to the Lie group $\rmGL_1(\bbC)$ which is isomorphic to the
Lie group $\bbC^\times = \bbC \smallsetminus \{0\}$. The mappings
$C_{\pm \alpha}$ are just complex functions on $\bbR^2$. The Toda
equations under consideration have the form (\ref{e:2.16}) with 
$p = n$.

We have shown in the paper \cite{NirRaz08} that if the functions $C_{-\alpha}$
and $C_{+\alpha}$ have no zeros then the Toda equations (\ref{e:2.16}) 
are equivalent to the same equations, but where $C_{-\alpha} = C_-$
and $C_{+\alpha} = C_+$ for some functions $C_-$ and $C_+$ which have
no zeros. If these functions are real, then with the help of an appropriate
change of the coordinates $z^-$ and $z^+$ we can come to the Toda equations with
$C_{\pm \alpha}$ equal to a nonzero constant $m$. This system 
of equations gives the Toda equations associated with untwisted 
loop groups of general linear groups. In the paper \cite{NirRaz08} we investigated
the soliton solutions of the above Toda equations obtained 
by two different approaches, the Hirota's and rational dressing 
methods, and established explicit relationships between these methods.

\subsection{Second type: twisted loop Toda equations, 
odd-dimensional case} \label{s:3.2}

The abelian Toda equations of the other two types arise when we
use outer automorphisms of $\gothgl_n(\bbC)$. For the equations 
of the second type $n$ is odd, and for the equations of the third 
type $n$ is even.

Consider first the case of an odd $n = 2s - 1$, $s \ge 2$. In this
case an abelian Toda equation arises when the automorphism $A$ is
defined by the equality
\begin{equation}
A(x) = - h (B^{-1} {}^{t\!} x B) h^{-1}, \label{e:3.2}
\end{equation}
where ${}^{t\!} x$ means the transpose of $x$, $h$ is a diagonal
matrix with the diagonal matrix elements
\[
h_{kk} = \epsilon_{2n}^{n-k+1} = \epsilon_{4s-2}^{2s-k},
\]
and $B$ is an $n \times n$ matrix of the form
\[
\psset{xunit=1.6em, yunit=1em}
B = \left( \raise -3.3\psyunit \hbox{\begin{pspicture}(.7,.5)(7.3,7.1)
\rput(1,7){$1$} \rput(7,6){$1$}
\qdisk(6.3,5.3){.7pt} \qdisk(6,5){.7pt} \qdisk(5.7,4.7){.7pt}
\rput(5,4){$1$} \rput(4,3){$-1$}
\qdisk(3.3,2.3){.7pt} \qdisk(3,2){.7pt} \qdisk(2.7,1.7){.7pt}
\rput(2,1){$-1$}
\psline(.7,6.5)(7.3,6.5)
\psline(1.5,7.3)(1.5,.5)
\end{pspicture}} \right).
\]
The corresponding group automorphism $a$ is defined as
\begin{equation}
a(g) = h (B^{-1} \: {}^{t\!}g^{-1} \: B) h^{-1}. \label{e:r6}
\end{equation}
The order $M$ of the automorphism $A$ is $2N = 4s-2$ and the integer 
$p$ is $2s - 1$. The mapping $\gamma$ is a diagonal matrix of the form 
(\ref{e:2.12}), where the mappings $\Gamma_\alpha$ are mappings of 
$\bbR^2$ to $\bbC^\times$, subject to the constraints
\[
\Gamma_1 = 1, \qquad \Gamma_{2s-\alpha+1} = \Gamma_\alpha^{-1}, \quad
\alpha = 2, \ldots, s.
\]
The mappings $C_{\pm \alpha}$ in the relations (\ref{e:2.13}) 
and (\ref{e:2.14}) are complex functions satisfying the equality
\begin{equation}
C_{\pm 0} = C_{\pm 1}, \label{e:3.3}
\end{equation}
and for $s > 2$ the equalities
\begin{equation}
C_{\pm (2s - \alpha)} = - C_{\pm \alpha}, \quad 
\alpha = 2, \ldots, s-1. \label{e:3.4}
\end{equation}
Let us choose the mappings $\Gamma_\alpha$, $\alpha = 2, \ldots, s$,
as a complete set of mappings parameterizing the mapping $\gamma$.
Taking into account the equalities (\ref{e:3.3}) and (\ref{e:3.4})
we come to a set of $s-1$ independent equations equivalent to
the Toda equation under consideration. As well as in the untwisted
case, under appropriate conditions the Toda equations under consideration are
equivalent to the same equations, but where
\begin{equation}
C_{\pm 0} = C_{\pm 1} = C_{\pm s} = m, \label{e:3.5}
\end{equation}
and 
\begin{equation}
C_{\pm \alpha} = - C_{\pm (2s-\alpha)} = m, \qquad 
\alpha = 2,\ldots,s-1. \label{e:3.6}
\end{equation}
Explicitly, we have the equations
\begin{align}
\partial_+(\Gamma_2^{-1} \partial_- \Gamma_2^{\ph}) & = - m^2
(\Gamma_2^{-1} \Gamma_3^{\ph} - \Gamma_2^{\ph}), \notag \\
\partial_+(\Gamma_3^{-1} \partial_- \Gamma_3^{\ph}) & = - m^2 (
\Gamma_3^{-1} \Gamma_4^{\ph} - \Gamma_2^{-1} \Gamma_3^{\ph}), 
\notag \\
& \vdots \label{e:3.7} \\
\partial_+(\Gamma_{s-1}^{-1} \partial_- \Gamma_{s-1}^{\ph}) & = - m^2
(\Gamma_{s-1}^{-1} \Gamma_s^{\ph} - \Gamma_{s-2}^{-1} \Gamma_{s-1}
^{\ph}), \notag \\
\partial_+(\Gamma_s^{-1} \partial_- \Gamma_s^{\ph}) & = - m^2
(\Gamma_s^{-2} - \Gamma_{s-1}^{-1} \Gamma_s^{\ph}), \notag
\end{align}
where $m$ is again a nonzero constant, see also the papers
\cite{Mik81,MikOlsPer81}.

For $s = 2$ denoting $\Gamma_2$ by $\Gamma$ we have the equation
\[
\partial_+(\Gamma^{-1} \partial_- \Gamma^{\ph}) = - m^2 (\Gamma^{-2} -
\Gamma^{\ph}).
\]
Putting $\Gamma = \exp(F)$ we obtain
\[
\partial_+ \partial_- F = - m^2 [\exp(-2F) -  \exp(F)].
\]
This is the well-known Dodd--Bullough--Mikhailov equation 
\cite{DodBul77,Mik81}, formulated for the first time by 
Tzitz\'eica \cite{Tzi08} in geometry of hyperbolic surfaces.

\subsection{Third type: twisted loop Toda equations, even-dimensional case} 
\label{s:3.3}

In the case of an even $n = 2s$, $s \ge 2$, to come to an abelian 
Toda equation we should use again the Lie algebra automorphism $A$
and the corresponding group automorphism $a$ defined by the relations
(\ref{e:3.2}) and (\ref{e:r6}), respectively, where now
\[
\psset{xunit=1.6em, yunit=1em}
B = \left( \raise -3.8\psyunit \hbox{\begin{pspicture}(-.3,.5)
(7.3,8.1)
\rput(1,8){$1$} \rput(0,7){$1$} \rput(7,6){$1$}
\qdisk(6.3,5.3){.7pt} \qdisk(6,5){.7pt} \qdisk(5.7,4.7){.7pt}
\rput(5,4){$1$} \rput(4,3){$-1$}
\qdisk(3.3,2.3){.7pt} \qdisk(3,2){.7pt} \qdisk(2.7,1.7){.7pt}
\rput(2,1){$-1$}
\psline(-.3,6.5)(7.3,6.5)
\psline(1.5,8.3)(1.5,.5)
\end{pspicture}} \right)
\]
and $h$ is a diagonal matrix with the diagonal matrix elements
\[
h_{11} = \epsilon_{2n-2}^{n-1} = \epsilon_{4s-2}^{2s-1} = -1, 
\qquad
h_{ii}= \epsilon_{2n-2}^{n-i+1} = \epsilon_{4s-2}^{2s-i+1}, 
\quad i = 2, \ldots, n.
\]
The order $M$ of the automorphism $A$ is again $2N = 4s - 2$, 
and the number $p$ characterizing the block structure is equal 
to $n-1 = 2s-1$, $n_1 = 2$, and $n_\alpha = 1$ for 
$\alpha = 2, \ldots, 2s-1$.

The mapping $\Gamma_1$ is a mapping of $\bbR^2$ to the Lie group
$\rmSO_2(\bbC)$ which is isomorphic to $\bbC^\times$. Actually
$\Gamma_1$ is a $2 \times 2$ complex matrix-valued function 
satisfying the relation
\[
J_2^{-1} \: {}^{t\!} \Gamma_1^\ph \: J_2^\ph = \Gamma_1^{-1},
\]
where
\[
J_2 = \left( \begin{array}{cc}
0 & 1 \\ 1 & 0
\end{array} \right).
\]
It is easy to show that $\Gamma_1$ has the form
\[
\Gamma_1 = \left( \begin{array}{cc}
(\Gamma_1)_{11}^\ph & 0 \\ 0 & (\Gamma_1)_{11}^{-1}
\end{array} \right),
\]
where $(\Gamma_1)_{11}$ is a mapping of $\bbR^2$ to $\bbC^\times$. The
mappings $\Gamma_\alpha$, $\alpha = 2, \ldots, 2s-1$, are mappings of
$\bbR^2$ to $\bbC^\times$ satisfying the relations
\[
\Gamma_{2s-\alpha+1}^\ph = \Gamma^{-1}_\alpha.
\]

The mappings $C_{-1}$, $C_{+0}$ are complex $1 \times 2$ matrix-valued
functions, the mappings $C_{-0}$, $C_{+1}$ are complex $2 \times 1$
matrix-valued functions. Here we have
\begin{equation}
C_{-0} = J^{-1}_2 {}^{t\!} C_{-1}, \qquad C_{+0} = {}^{t\!} C_{+1}
J_2^\ph. \label{e:3.8}
\end{equation}
The mappings $C_{\pm \alpha}$, $\alpha = 2, \ldots, p - 1 = 2s - 2$,
are just complex functions, satisfying for $s > 2$ the equalities
\begin{equation}
C_{\pm(2s - \alpha)} = -C_{\pm \alpha}, \qquad \alpha = 2, \ldots,
s-1. \label{e:3.9}
\end{equation}

The mappings $(\Gamma_1)_{11}$ and $\Gamma_\alpha$, $\alpha = 2,
\ldots, s$, form a complete set of mappings parameterizing the 
mapping $\gamma$. Taking into account the equalities (\ref{e:3.8}) 
and (\ref{e:3.9}) we come to a set of $s$ independent equations
equivalent to the Toda equation under consideration. As well as 
for the first two types, under appropriate conditions these
equations can be reduced to equations with
\begin{equation}
C_{-\alpha} = m, \qquad C_{+\alpha} = m, \qquad \alpha = 2,\ldots,s,
\label{e:3.10}
\end{equation}
and
\begin{equation}
(C_{-1})_{11} = (C_{-1})_{12} = m / \sqrt{2}, \qquad 
(C_{+1})_{11} = (C_{+1})_{21} = m / \sqrt{2}, 
\label{e:3.11}
\end{equation}
where $m$ is a nonzero constant. Thus, we come to the equations
\begin{align}
\partial_+(\Gamma_1^{-1} \partial_- \Gamma_1^\ph) = & - \frac{m^2}{2}
(\Gamma_1^{-1} - \Gamma_1^\ph) \Gamma_2^{\ph}, \notag \\
\partial_+(\Gamma_2^{-1} \partial_- \Gamma_2^{\ph}) = & - m^2
\Gamma_2^{-1} \Gamma_3^{\ph} + \frac{m^2}{2} (\Gamma_1^{-1} +
\Gamma_1^{\ph})\Gamma_2^{\ph}, \notag \\
\partial_+(\Gamma_3^{-1} \partial_- \Gamma_3^{\ph}) = & - m^2
(\Gamma_3^{-1} \Gamma_4^{\ph} - \Gamma_2^{-1} \Gamma_3^{\ph}), 
\notag \\
& \vdots \label{e:3.12} \\
\partial_+(\Gamma_{s-1}^{-1} \partial_- \Gamma_{s-1}^{\ph}) = & - m^2
(\Gamma_{s-1}^{-1} \Gamma_s^{\ph} - \Gamma_{s-2}^{-1} \Gamma_{s-1}
^{\ph}), \notag \\
\partial_+(\Gamma_s^{-1} \partial_- \Gamma_s^{\ph}) = & - m^2
(\Gamma_s^{-2} - \Gamma_{s-1}^{-1} \Gamma_s^{\ph}), \notag
\end{align}
where, with a slight abuse of notation, we have denoted 
$(\Gamma_1)_{11}$ by $\Gamma_1$.

We also note that all three systems of Toda equations described above
can be represented in standard forms with explicit indication of the
Cartan matrices of the corresponding affine Lie algebras of the types
$A^{(1)}_{n-1}$, $A^{(2)}_{2s-2}$ and $A^{(2)}_{2s-1}$, respectively, see,
for example, the paper \cite{NirRaz08}.

\section{Rational dressing} \label{s:4}

In this section we apply the method of rational dressing to construct 
solutions of the abelian Toda systems associated with the loop groups 
of the complex general linear groups. Here we solve the abelian Toda 
equations of the second and third types which have the forms (\ref{e:3.7}) 
and (\ref{e:3.12})respectively. In fact, some preliminary 
relations of the rational dressing formalism can be introduced on a 
common basis in application to the both types of abelian Toda systems. 

Because in the cases under consideration the matrices $c_-$ and $c_+$ 
are commuting, it is obvious that
\begin{equation}
\gamma = I_n,
\label{e:4.1}
\end{equation}
where $I_n$ is the $n \times n$ unit matrix, is a solution to the Toda
equation (\ref{e:2.7}). Denote a mapping of $\bbR \times S^1$ to
$\rmGL_n(\bbC)$, which generates the corresponding connection, by
$\varphi$. Using the equalities (\ref{e:2.5}) and (\ref{e:2.6}) 
and remembering that in our case $L = 1$, we write
\begin{equation}
\varphi^{-1} \partial_- \varphi = \lambda^{-1} c_-, \qquad
\varphi^{-1} \partial_+ \varphi = \lambda \: c_+,
\label{e:r2}
\end{equation}
where the matrices $c_+$ and $c_-$ having generally the forms 
(\ref{e:2.13}) and (\ref{e:2.14}), are specified by the relations
(\ref{e:3.5}), (\ref{e:3.6}) for the Toda equations of the second
type, and by the relations (\ref{e:3.10}), (\ref{e:3.11}) 
for the Toda equations of the third type.

To construct some other solutions to the Toda equations we will look for 
a mapping $\psi$, such that the mapping
\begin{equation}
\varphi' = \varphi \: \psi
\label{e:4.2}
\end{equation}
would generate a connection satisfying the grading condition 
\begin{equation}
\omega_- = \omega_{-0} + \omega_{-1}, \qquad 
\omega_+ = \omega_{+0} + \omega_{+1} \label{e:r3}
\end{equation}
and the gauge-fixing constraint 
\begin{equation}
\omega_{+0} = 0. \label{e:r4}
\end{equation}

For any $m \in \bbR^2$ the mapping $\tilde \psi_m$ defined by the
equality $\tilde \psi_m(p) = \psi(m, p)$, $p \in S^1$, is a smooth
mapping of $S^1$ to $\rmGL_n(\bbC)$. We treat $S^1$ as a subset of 
the complex plane which, in turn, will be treated as a subset of the 
Riemann sphere. Assume that it is possible to extend analytically 
each mapping $\tilde \psi_m$ to all of the Riemann sphere. As the 
result we obtain a mapping of the direct product of $\bbR^2$ and 
the Riemann sphere to $\rmGL_n(\bbC)$, which we also denote by 
$\psi$. Suppose that for any $m \in \bbR^2$ the analytic extension 
of $\tilde \psi_m$ results in a rational mapping regular at the 
points $0$ and $\infty$, hence the name {\em rational dressing\/}. 
Below, for each point $p$ of the Riemann sphere we denote by 
$\psi_p$ the mapping of $\bbR^2$ to $\rmGL_n(\bbC)$ defined by 
the equality $\psi_p(m) = \psi(m, p)$.

Since we deal with the Toda equations described in Sections
\ref{s:3.2} and \ref{s:3.3}, for any $m \in \bbR^2$ and 
$p \in S^1$ we should have
\begin{equation}
\psi(m, \epsilon_{2N} p) 
= h \: B^{-1} \: {}^{t\!}\psi^{-1}(m, p) \: B \: h^{-1},
\label{e:4.3}
\end{equation}
where $h$ is a block-diagonal matrix described by the relation
\[
h_{\alpha\beta} = \epsilon^{N-\alpha+1}_{2N} I_{n_\alpha} 
\delta_{\alpha\beta}, \qquad \alpha,\beta = 1,\ldots,p,
\]
with $n_1 = 1$ for the Toda equations of the second type, and
$n_1=2$ for the Toda equations of the third type, while for all
other indices $\alpha = 2,\ldots,p$ we always have $n_\alpha = 1$. 
Note that $h_{11} = -I_{n_1}$. Here we also use the notation
\[
\psset{xunit=1.6em, yunit=1em}
B = \left( \raise -3.8\psyunit \hbox{\begin{pspicture}(-.3,.5)
(7.3,8.1)
\rput(0.5,7.5){$J_{n_1}$} \rput(7,6){$1$}
\qdisk(6.3,5.3){.7pt} \qdisk(6,5){.7pt} \qdisk(5.7,4.7){.7pt}
\rput(5,4){$1$} \rput(4,3){$-1$}
\qdisk(3.3,2.3){.7pt} \qdisk(3,2){.7pt} \qdisk(2.7,1.7){.7pt}
\rput(2,1){$-1$}
\psline(-.3,6.5)(7.3,6.5)
\psline(1.5,8.3)(1.5,.5)
\end{pspicture}} \right)
\]  
common for the both cases. The equality (\ref{e:4.3}) means that
for any $m \in \bbR^2$ two rational mappings coincide on $S^1$,
therefore, they must coincide on the entire Riemann sphere.

We define a linear mapping $\hat a$ acting on a mapping $\chi$
of the direct product of $\bbR^2$ and the Riemann sphere to the 
algebra $\Mat_n(\bbC)$ of $n \times n$ complex matrices 
as\footnote{Note that below $\chi$ is a mapping to the Lie group
$\rmGL_n(\bbC)$, although to justify the relation (\ref{e:r.1})
it is convenient to think $\rmGL_n(\bbC)$ as a
subset of $\Mat_n(\bbC)$.} 
\[
\hat a \chi (m,p) 
= h B^{-1} {}^t \chi^{-1} (m, \epsilon_{2N}^{-1} p) B h^{-1}.
\]
The relation (\ref{e:4.3}) is equivalent to the equality 
$\hat a \psi = \psi$. To construct rational mappings 
satisfying this relation we will use the following
procedure. First, we construct a family of mappings 
$\psi$ satisfying the relation $\hat a^2 \psi = \psi$, 
and then select from it the mappings satisfying
the equality $\hat a \psi = \psi$.

It is easy to see that the mapping
\begin{equation}
\psi = \sum_{k=1}^N \hat a^{2k} \chi \label{e:r.1}
\end{equation}
satisfies the relation $\hat a^2 \: \psi = \psi$. It is worth 
to note that $\hat a^{2N} \chi = \chi$. We start with a rational 
mapping $\chi$ regular at the points $0$ and $\infty$ and having 
poles at $r$ different nonzero points $\mu_i$, $i = 1, \ldots, r$. 
More specifically, we consider a mapping $\chi$ of the form
\[
\chi = \chi_0 \left( I_n + \sum_{i=1}^r 
\frac{\lambda}{\lambda - \mu_i} P_i \right),
\]
where $P_i$ are some smooth mappings of $\bbR^2$ to the algebra 
$\Mat_n(\bbC)$ and $\chi_0$ is a mapping of $\bbR^2$ to the Lie 
subgroup of $\rmGL_n(\bbC)$ formed by the elements 
$g \in \rmGL_n(\bbC)$ satisfying the equality
\begin{equation}
h^2 g h^{-2} = g. \label{e:4.5}
\end{equation}

With account of the equality
\[
\hat a^2 \: \chi (m, p) = h^2 \: \chi(m, \epsilon_N^{-1} p) \: h^{-2}
\]
the averaging procedure (\ref{e:r.1}) leads to the mapping
\begin{equation}
\psi = \psi_0 \left(I_n + \sum_{i=1}^r \sum_{k=1}^N \frac{\lambda}
{\lambda - \epsilon_{2N}^{2k} \mu_i} h^{2k} P_i \: h^{-2k} \right),
\label{e:4.6}
\end{equation}
where $\psi_0 = N \chi_0$. We assume that 
$\mu_i^{2N} \ne \mu_j^{2N}$ for all $i \ne j$.

Denote by $\psi^{-1}$ the mapping of $\bbR^2 \times S^1$ to
$\rmGL_n(\bbC)$ defined by the relation
\[
\psi^{-1}(m, p) = (\psi(m, p))^{-1}.
\]
Suppose that for any fixed $m \in \bbR^2$ the mapping
$\tilde\psi_m^{-1}$ of $S^1$ to $\rmGL_n(\bbC)$, defined by the
equality $\tilde\psi_m^{-1}(p) = \psi^{-1}(m,p)$, can be extended
analytically to a mapping of the Riemann sphere to $\rmGL_n(\bbC)$,
which we also denote by $\psi^{-1}$, and as the result we obtain
a rational mapping of the same structure as the mapping $\psi$,
\begin{equation}
\psi^{-1} = \left( I_n + \sum_{i = 1}^r \sum_{k=1}^N \frac{\lambda}
{\lambda - \epsilon_{2N}^{2k} \nu_i} h^{2k} Q_i h^{-2k} 
\right) \psi_0^{-1},
\label{e:4.7}
\end{equation}
with the pole positions satisfying the conditions $\nu_i \ne 0$,
$\nu_i^{2N} \ne \nu_j^{2N}$ for all $i \ne j$, and additionally 
$\nu_i^{N} \ne \mu_j^{N}$ for any $i$ and $j$.
\footnote{Actually, as it will be clear, for the extended mappings
$\psi$ and $\psi^{-1}$ we have $\psi^{-1} \psi = I_n$. This justifies
the notation used.} 

The mappings $\psi$ and $\psi^{-1}$ given by the equalities (\ref{e:4.6})
and (\ref{e:4.7}), respectively, satisfy the relations 
$\hat a^2 \psi = \psi$ and $\hat a^2 \psi^{-1} = \psi^{-1}$. To
satisfy the relations $\hat a \psi = \psi$ and 
$\hat a \psi^{-1} = \psi^{-1}$ we have to assume
that the pole positions of the mappings $\psi$ and 
$\psi^{-1}$ are necessarily connected as
\[
\nu_i = \mu_i / \epsilon_{2N}, \qquad i = 1,\ldots,r,
\]
and the matrices $P_i$ and $Q_i$ are related as
\begin{equation}
Q_i = h^{-1} \: B^{-1} \: {}^{t\!} P_i \: B \: h, \qquad 
i = 1,\ldots,r,
\label{e:4.9}
\end{equation}

By definition, the equality
\[
\psi^{-1} \psi = I_n
\]
is valid at all points of the direct product of $\bbR^2$ and $S^1$.
Since $\psi^{-1} \psi$ is a rational mapping, the above equality is
valid at all points of the direct product of $\bbR^2$ and the Riemann
sphere. Hence, the residues of $\psi^{-1} \psi$ at the points 
$\nu_i = \mu_i/\epsilon_{2N}$ and $\mu_i$ should be equal to zero. 
Explicitly we have
\begin{gather}
h^{-1} B^{-1} {}^{t\!}P_i B h 
\left( I_n + \sum_{j = 1}^r \sum_{k=1}^N \frac{\mu_i/\epsilon_{2N}}
{\mu_i/\epsilon_{2N} - \epsilon_{2N}^{2k} \mu_j} h^{2k} P_j h^{-2k} 
\right) = 0,
\label{e:4.10} \\
\left( I_n + \sum_{j = 1}^r \sum_{k=1}^N \frac{\mu_i}{\mu_i -
\epsilon_{2N}^{2k-1} \mu_j} h^{2k-1} B^{-1} {}^{t\!}P_j B h^{-2k+1}
\right) P_i = 0.
\label{e:4.11}
\end{gather}
We will discuss later how to satisfy these relations, and now let 
us consider what connection is generated by the mapping $\varphi'$ 
defined by (\ref{e:4.2}) with the mapping $\psi$ possessing the 
properties described above.

Using the equality (\ref{e:4.2}) and the relations (\ref{e:r2}), 
we obtain for the components of the connection generated by 
$\varphi'$ the expressions
\begin{gather}
\omega_- = \psi^{-1} \partial_- \psi + \lambda^{-1} \psi^{-1} c_-
\psi, \label{e:4.12} \\
\omega_+ = \psi^{-1} \partial_+ \psi + \lambda \psi^{-1} c_+ \psi.
\label{e:4.13}
\end{gather}
We see that the component $\omega_-$ is a rational mapping which has
simple poles at the points $\mu_i$, $\nu_i=\mu_i/\epsilon_{2N}$ and 
zero.\footnote{Here and below discussing the holomorphic properties 
of mappings and functions we assume that the point of the space 
$\bbR^2$ is arbitrary but fixed.} Similarly, the component $\omega_+$ 
is a rational mapping which has simple poles at the points $\mu_i$, 
$\nu_i=\mu_i/\epsilon_{2N}$ and infinity. We are looking for a 
connection which satisfies the grading condition (\ref{e:r3})
and the gauge-fixing condition (\ref{e:r4}). 
The grading condition in our case is the requirement that for each 
point of $\bbR^2$ the component $\omega_-$ is rational and has the 
only simple pole at zero, while the component $\omega_+$ is rational 
and has the only simple pole at infinity. Hence, we demand that the 
residues of $\omega_-$ and $\omega_+$ at the points $\mu_i$ and 
$\nu_i=\mu_i/\epsilon_{2N}$ should vanish.

The residues of $\omega_-$ and $\omega_+$ at the points 
$\nu_i=\mu_i/\epsilon_{2N}$ are equal to zero if and only if
\begin{gather}
(\partial_- Q_i - \epsilon_{2N}\mu_i^{-1} Q_i c_-) 
\left( I_n + \sum_{j = 1}^r \sum_{k=1}^N 
\frac{\mu_i/\epsilon_{2N}}{\mu_i/\epsilon_{2N} 
- \epsilon_{2N}^{2k} \mu_j} h^{2k} P_j h^{-2k}
\right) = 0,
\label{e:4.14} \\
(\partial_+ Q_i - \epsilon^{-1}_{2N} \mu_i Q_i c_+) 
\left( I_n + \sum_{j = 1}^r 
\sum_{k=1}^N \frac{\mu_i/\epsilon_{2N}}{\mu_i/\epsilon_{2N} 
- \epsilon_{2N}^{2k} \mu_j} h^{2k} P_j h^{-2k} 
\right) = 0,
\label{e:4.15}
\end{gather}
respectively, with the equality (\ref{e:4.9}) to be taken into account. 
Similarly, the requirement of vanishing of the residues at 
the points $\mu_i$ gives the relations
\begin{gather}
\left( I_n + \sum_{j = 1}^r \sum_{k=1}^N \frac{\mu_i}{\mu_i -
\epsilon_{2N}^{2k-1} \mu_j} h^{2k-1} B^{-1} {}^{t\!}P_j B h^{-2k+1} 
\right) (\partial_- P_i + \mu_i^{-1} c_- P_i) = 0,
\label{e:4.16} \\
\left( I_n + \sum_{j = 1}^r \sum_{k=1}^N \frac{\mu_i}{\mu_i -
\epsilon_{2N}^{2k-1} \mu_j} h^{2k-1} B^{-1} {}^{t\!}P_j B h^{-2k+1} 
\right) (\partial_+ P_i + \mu_i c_+ P_i) = 0.
\label{e:4.17}
\end{gather}
To obtain the relations (\ref{e:4.14})--(\ref{e:4.17}) we made use 
of the equalities (\ref{e:4.10}), (\ref{e:4.11}).

Suppose that we have succeeded in satisfying the relations 
(\ref{e:4.10}), (\ref{e:4.11}) and (\ref{e:4.14})--(\ref{e:4.17}). 
In such a case from the equalities (\ref{e:4.12}) and (\ref{e:4.13}) 
it follows that the connection under consideration satisfies the 
grading condition.

It follows from the equality 
(\ref{e:4.13}) that
\[
\omega_+(m, 0) = \psi_0^{-1}(m) \partial_+ \psi_0(m).
\]
Taking into account that $\omega_{+0}(m) = \omega_+(m, 0)$, we
conclude that the gauge-fixing constraint $\omega_{+0} = 0$ is
equivalent to the relation
\begin{equation}
\partial_+ \psi_0 = 0. \label{e:4.18}
\end{equation}
Assuming that this relation is satisfied, we come to a connection
satisfying both the grading condition and the gauge-fixing condition.

Recall that if a flat connection $\omega$ satisfies the grading and
gauge-fixing conditions, then there exist a mapping $\gamma$ from
$\bbR^2$ to $G$ and mappings $c_-$ and $c_+$ of $\bbR^2$ to
$\gothg_{-1}$ and $\gothg_{+1}$, respectively, such that the
representation (\ref{e:2.6}) for the components $\omega_-$  and
$\omega_+$ is valid. In general, the mappings $c_-$ and $c_+$
parameterizing the connection components may be different from the
mappings $c_-$ and $c_+$ which determine the mapping $\varphi$. Let us
denote the mappings corresponding to the connection under
consideration by $\gamma'$, $c_-'$ and $c_+'$. Thus, we have
\begin{align}
\psi^{-1} \partial_- \psi + \lambda^{-1} \psi^{-1} c_- \psi &=
\gamma^{\prime -1} \partial_- \gamma' + \lambda^{-1} c_-',
\label{e:4.19} \\
\psi^{-1} \partial_+ \psi + \lambda \psi^{-1} c_+ \psi &= \lambda
\gamma^{\prime -1} c_+' \gamma'. \label{e:4.20}
\end{align}
Note that $\psi_\infty$ is a mapping of $\bbR^2$ to the Lie subgroup
of $\rmGL_n(\bbC)$ defined by the relation (\ref{e:4.5}). We recall that
this subgroup coincides with $G_0$ and denote $\psi_\infty$ by
$\gamma$. From the relation (\ref{e:4.19}) we obtain the equality
\[
\gamma^{\prime -1} \partial_- \gamma' 
= \gamma^{-1} \partial_- \gamma.
\]
The same relation (\ref{e:4.19}) gives
\[
\psi^{-1}_0 c_- \psi_0 = c_-'.
\]
Impose the condition $\psi_0 = I_n$, which is consistent with
the condition (\ref{e:4.18}). Here we have
\[
c_-' = c_-.
\]
Finally, from the equality (\ref{e:4.20}) we obtain
\[
\gamma^{\prime -1} c_+' \gamma' = \gamma^{-1} c_+ \gamma.
\]
We see that if we impose the condition $\psi_0 = I_n,$ then the
components of the connection under consideration have the form 
(\ref{e:2.6}) where $\gamma = \psi_\infty$.

Thus, to find solutions to the Toda equations under
consideration, we can use the following procedure. We fix 
$2r$ complex numbers $\mu_i$ and $\nu_i$ and find matrix-valued 
functions $P_i$ and $Q_i$ satisfying the relations 
(\ref{e:4.10}), (\ref{e:4.11}) and
(\ref{e:4.14})--(\ref{e:4.17}). With the help 
of the relations (\ref{e:4.6}), (\ref{e:4.7}), 
assuming that
\[
\psi_0 = I_n,
\]
we construct the mappings $\psi$ and $\psi^{-1}$. Then, the mapping
\begin{equation}
\gamma = \psi_\infty
\label{e:4.21}
\end{equation}
satisfies the Toda equation (\ref{e:2.7}).

Let us return to the relations (\ref{e:4.10}), (\ref{e:4.11}).
It is easy to see that they are equivalent, and so, we will use the 
relation (\ref{e:4.11}) for further calculations. We can show that, 
if we suppose that the matrix $P_i$ has the maximum rank, then we 
get the trivial solution of the Toda equation given by (\ref{e:4.1}). 
Hence, we will assume that $P_i$ is not of maximum rank. The simplest 
case here is given by matrices of rank one which can be represented as
\[
P_i = u^{}_i {}^{t\!} w^{}_i, 
\]
where $u$ and $w$ are $n$-dimensional column vectors. This
representation allows writing the relations (\ref{e:4.11}) 
as
\begin{equation}
u^{}_i + \sum_{j=1}^r \sum_{k=1}^N \frac{\mu_i}{\mu_i 
- \epsilon_{2N}^{2k-1} \mu_j} h^{2k-1} B^{-1} w^{}_j 
({}^{t\!} u_j B h^{-2k+1} u^{}_i) = 0. 
\label{e:4.22}
\end{equation}
Using the identity
\[
\sum_{k=1}^{N} \frac{z \epsilon_{2N}^{-2kj}}
{z - \epsilon_{2N}^{2k}} = N \frac{z^{N-|j|_N}}{z^N - 1},
\]
where $|j|_N$ is the residue of division of $j$ by $N$, 
we can rewrite the equality (\ref{e:4.22}) in terms of the components 
of $u_i$ as follows:
\[
{}^{t\!}u_{i,1} J_{n_1} + N \sum_{j=1}^r 
(R_1)_{i j} {}^{t\!}w_{j,1} = 0,
\]
where $u_{i,1}$ and $w_{i,1}$ gather first $n_1$ components of the
corresponding $n$-dimensional column vectors, so these are in fact
$n_1$-dimensional column vectors,\footnote{We remember that either 
$n_1 = 1$ or $n_1 = 2$.}
\[
u_{i,N+2-k} - N \sum_{j=1}^r 
(R_k)_{i j} w_{j,k} = 0, \qquad k = 2,\ldots,s,
\]
and
\[
u_{i,N+2-k} + N \sum_{j=1}^r 
(R_k)_{i j} w_{j,k} = 0, \qquad k = s+1,\ldots,p=N.
\]
Here the $r \times r$ matrices $R_1$ and $R_k$ are defined as
\begin{eqnarray}
(R_1)_{i j} &=& \frac{1}{\mu_i^N + \mu_j^N} 
\left( \mu_i^N ({}^{t\!}u_{i,1} J_{n_1} u_{j,1}) - \sum_{\ell=2}^s 
\mu_i^{N-|\ell-1|_N} \mu_j^{|\ell-1|_N} (u_{i,N+2-\ell} u_{j,\ell}) 
\right. \notag \\* 
&& \left. \hskip4.5cm   
+ \sum_{\ell=s+1}^N \mu_i^{N-|\ell-1|_N} \mu_j^{|\ell-1|_N}
(u_{i,N+2-\ell} u_{j,\ell}) \right), \label{e:4.23} \\
(R_k)_{i j} &=& \frac{1}{\mu_i^N + \mu_j^N} 
\Biggl( -\mu_i^{N-|1-k|_N} \mu_j^{|1-k|_N}
({}^{t\!}u_{i,1} J_{n_1} u_{j,1}) \notag \\*
&& + \sum_{\ell=2}^{k-1}
\mu_i^{N-|\ell-k|_N} \mu_j^{|\ell-k|_N} (u_{i,N+2-\ell} u_{j,\ell})
- \sum_{\ell=k}^s \mu_i^{N-|\ell-k|_N} \mu_j^{|\ell-k|_N}
(u_{i,N+2-\ell} u_{j,\ell}) \notag \\*  
&& \hskip4.5cm 
+ \sum_{\ell=s+1}^N \mu_i^{N-|\ell-k|_N} \mu_j^{|\ell-k|_N}
(u_{i,N+2-\ell} u_{j,\ell}) \Biggr) \label{e:4.24} 
\end{eqnarray}
for $k = 2,\ldots,s$, and
\begin{eqnarray}
(R_k)_{i j} &=& \frac{1}{\mu_i^N + \mu_j^N} 
\Biggl( -\mu_i^{N-|1-k|_N} \mu_j^{|1-k|_N}
({}^{t\!}u_{i,1} J_{n_1} u_{j,1}) \notag \\* 
&& + \sum_{\ell=2}^{s}
\mu_i^{N-|\ell-k|_N} \mu_j^{|\ell-k|_N} (u_{i,N+2-\ell} u_{j,\ell})
- \sum_{\ell=s+1}^{k-1} \mu_i^{N-|\ell-k|_N} \mu_j^{|\ell-k|_N}
(u_{i,N+2-\ell} u_{j,\ell}) \notag \\* 
&& \hskip5cm 
+ \sum_{\ell=k}^N \mu_i^{N-|\ell-k|_N} \mu_j^{|\ell-k|_N}
(u_{i,N+2-\ell} u_{j,\ell}) \Biggr)
\label{e:4.25}
\end{eqnarray}
for $k=s+1,\ldots,N$. Recall that for all cases considered here
$N=p=2s-1$. 

We use the equations (\ref{e:4.23}), (\ref{e:4.24}) and (\ref{e:4.25}) 
to express the vectors $w_i$ via the vectors $u_i$,
\[
{}^{t\!}w_{i,1} = - \frac{1}{N} \sum_{j=1}^r (R^{-1}_1)_{i j} 
{}^{t\!}u_{j,1} J_{n_1}, \qquad
w_{i,k} = \frac{1}{N} \sum_{j=1}^r (R^{-1}_k)_{i j} u_{j,N+2-k} 
\]
for $k=2,\ldots,s$, and
\[
w_{i,k} = - \frac{1}{N} \sum_{j=1}^r (R^{-1}_k)_{i j} u_{j,N+2-k}
\]
for $k=s+1,\ldots,N=p$.
As the result, having expressed the matrices $P_i$ and $Q_i$
in terms of the components of the vectors $u_i$, we find a 
solution of the relations (\ref{e:4.10}) and (\ref{e:4.11}). 

Further, it follows from the equality (\ref{e:4.22}) that, 
to fulfill also the relations (\ref{e:4.14})--(\ref{e:4.17}), 
it is sufficient to satisfy the equations
\[
\partial_- u_i = - \mu_i^{-1} c_- u_i, \qquad 
\partial_+ u_i = - \mu_i^{} c_+ u_i. 
\]
The general solution to these equations is given formally by the
expression
\begin{equation}
u_i(z^-,z^+) = \exp(-\mu_i^{-1} c_- z^- -\mu_i^{} c_+ z^+) u^0_i,
\label{e:4.26}
\end{equation}
where $u^0_i = u_i(0,0)$. We will make explicit this formal solution
when later constructing soliton solutions. 

Thus, we see that it is possible to satisfy the relations 
(\ref{e:4.10}), (\ref{e:4.11}) and (\ref{e:4.14})--(\ref{e:4.17}). 
This gives us solutions of the Toda equation (\ref{e:2.7}), 
and so, to the equations (\ref{e:3.7}) and (\ref{e:3.12}) 
by specifying the above formal expression of $u_i$ for the two 
corresponding cases. Let us show that they can be written in a 
simple determinant form.

Using the equalities (\ref{e:4.21}) and (\ref{e:4.6}), we get
\[
\gamma = \psi_\infty = I_n + \sum_{i = 1}^r \sum_{k=1}^N h^{2k} 
\: P_i \: h^{-2k}.
\]
For the matrix elements of $\gamma$ this gives the expression
\[
\gamma_{k \ell} = \delta_{k \ell} 
\left( 1 + N \sum_{i = 1}^r (P_i)_{kk} \right)
= \delta_{k \ell} \: \Gamma_k.
\]
Hence, we have
\[
\Gamma_1 = I_{n_1} - \sum_{i,j=1}^r 
u_{i,1} \: (R^{-1}_1)_{ij} \: {}^{t\!}u_{j,1} \: J_{n_1}, \qquad
\Gamma_\alpha = 1 + \sum_{i,j = 1}^r u_{i,\alpha} 
\: (R^{-1}_\alpha)_{ij} \: u_{j,2s+1-\alpha},
\]
where $\alpha = 2, \ldots, s$, and
\[
\Gamma_\alpha = 1 - \sum_{i,j = 1}^r u_{i,\alpha} 
\: (R^{-1}_\alpha)_{ij} \: u_{j,2s+1-\alpha}, \qquad
\alpha = s+1,\ldots,2s-1.
\]

We assume for convenience that the functions $u_{i,\alpha}$ are 
defined for arbitrary integral values of $\alpha$ so that
\[
u_{i,2s - 1 + \alpha} = u_{i,\alpha}.
\]
By definition the matrices $R_\alpha$ are periodic in the 
index $\alpha$. It appears that it is more appropriate to use 
quasi-periodic quantities $\wt u_\alpha$ and $\wt R_\alpha$
defined as
\[
\wt u_\alpha = M^\alpha \: u_\alpha, \qquad 
 \wt R_1 = M^{} \: R_1 \: M^{2s}, \qquad
\wt R_\alpha = M^{2s+1-\alpha} \: R_\alpha \: M^\alpha,
\]
where $\alpha = 2,\ldots,2s-1$; here $M$ is a diagonal 
$r \times r$ matrix given by
\[
M_{i j} = \mu_i \delta_{i j}.
\]
For these quantities we have quasi-periodicity conditions
\[
\wt u_{2s-1+\alpha} = M^{2s-1} \: \wt u_\alpha, \qquad
\wt R_{2s} = M^{2s-1} \: \wt R_1 \: M^{2s-1}, \qquad
\wt R_{2s-1+\alpha} = M^{-2s+1} \: \wt R_\alpha \: M^{2s-1}.
\]
The expression of the matrix elements of the matrices $\wt R_\alpha$
through the quasi-periodic quantities $\wt u_{i \alpha}$
has a remarkably simple form. We have for $\alpha=1$
\begin{multline*}
(\wt R_1)_{i j} = \frac{1}{\mu_i^{2s-1} + \mu_j^{2s-1}} \left( 
\mu_i^{2s-1} ({}^{t\!}\wt u_{i,1} J_{n_1} \wt u_{j,1}) \mu_j^{2s-1} 
- \mu_j^{2s-1} \sum_{\beta=2}^{s} \wt u_{i,2s+1-\beta} \wt u_{j,\beta} 
\right. \\ \left. + \mu_j^{2s-1} \sum_{\beta=s+1}^{2s-1} 
\wt u_{i,2s+1-\beta} \wt u_{j,\beta} \right).
\end{multline*}
Further, we have for $\alpha = 2,\ldots,s$
\begin{multline*}
(\wt R_\alpha)_{i j} = \frac{1}{\mu_i^{2s-1} + \mu_j^{2s-1}} \left( 
-\mu_i^{2s-1} ({}^{t\!}\wt u_{i,1} J_{n_1} \wt u_{j,1}) \mu_j^{2s-1} 
+ \mu_j^{2s-1} \sum_{\beta=2}^{\alpha-1} \wt u_{i,2s+1-\beta} \wt u_{j,\beta} 
\right. \\ \left. - \mu_i^{2s-1} \sum_{\beta=\alpha}^{s} 
\wt u_{i,2s+1-\beta} \wt u_{j,\beta} + \mu_i^{2s-1} \sum_{\beta=s+1}^{2s-1} 
\wt u_{i,2s+1-\beta} \wt u_{j,\beta} 
\right),  
\end{multline*}
and for $\alpha = s+1,\ldots,2s-1$
\begin{multline*}
(\wt R_\alpha)_{i j} = \frac{1}{\mu_i^{2s-1} + \mu_j^{2s-1}} \left( 
-\mu_i^{2s-1} ({}^{t\!}\wt u_{i,1} J_{n_1} \wt u_{j,1}) \mu_j^{2s-1} 
+ \mu_j^{2s-1} \sum_{\beta=2}^{s} \wt u_{i,2s+1-\beta} \wt u_{j,\beta} 
\right. \\ \left. - \mu_j^{2s-1} \sum_{\beta=s+1}^{\alpha-1} 
\wt u_{i,2s+1-\beta} \wt u_{j,\beta} 
+ \mu_i^{2s-1} \sum_{\beta=\alpha}^{2s-1} 
\wt u_{i,2s+1-\beta} \wt u_{j,\beta} 
\right).
\end{multline*}
Here we used the identity $|\!-k|_N = N - 1 - |k - 1|_N$.
The quasi-periodic functions have the following useful properties:
\begin{gather}
(\wt R_{\alpha + 1})_{i j} = (\wt R_{\alpha})_{i j} + 
\wt u_{i,2s+1-\alpha} \wt u_{j,\alpha}, \qquad
\alpha = 2,\ldots,s,
\label{e:4.27} \\
(\wt R_{\alpha + 1})_{i j} = (\wt R_{\alpha})_{i j} - 
\wt u_{i,2s+1-\alpha} \wt u_{j,\alpha}, \qquad
\alpha = s+1,\ldots,2s-1,
\label{e:4.28}
\end{gather}
and
\begin{equation}
(\wt R_1)_{i j} = - (\wt R_2)_{j i}, \qquad
(\wt R_\alpha)_{i j} = (\wt R_{2s + 2 - \alpha})_{j i}, \qquad
\alpha = 2,\ldots,2s-1.
\label{e:4.29}
\end{equation}
In terms of the quasi-periodic quantities, for the $n_1 \times n_1$
matrix-valued function $\Gamma_1$ and for the functions $\Gamma_\alpha$
we have
\[
\Gamma_1 = I_{n_1} - \sum_{i,j=1}^r \mu_i^{2s-1} \: \wt u_{i,1}
\: (\wt R_1^{-1})_{i j} \: {}^{t\!} \wt u_{j,1} \: J_{n_1}, \qquad
\Gamma_\alpha = 1 + \sum_{i,j=1}^r \wt u_{i,\alpha} 
\: (\wt R^{-1}_\alpha)_{i j} \: \wt u^{}_{j,2s+1-\alpha},
\]
for $\alpha = 2,\ldots,s$, and
\[
\Gamma_\alpha = 1 - \sum_{i,j=1}^r \wt u_{i,\alpha} 
\: (\wt R^{-1}_\alpha)_{i j} \: \wt u^{}_{j,2s+1-\alpha},
\]
for $\alpha = s+1,\ldots,2s-1$. The expressions for the functions
$\Gamma_\alpha$ for $\alpha > 1$ can be written as
\[
\Gamma_\alpha = 1 + {}^{t\!} \wt u_\alpha \: \wt R_\alpha^{-1} \: 
\wt u_{2s+1-\alpha}, \qquad \alpha = 2,\ldots,s, 
\]
and
\[
\Gamma_\alpha = 1 - {}^{t\!} \wt u_\alpha \: \wt R_\alpha^{-1} \: 
\wt u_{2s+1-\alpha}, \qquad \alpha = s+1,\ldots,2s-1. 
\]
Here $\wt R_\alpha$ is an $r \times r$ matrix and $\wt u_\alpha$
is an $r$-dimensional column vector. We remember that in the cases
under consideration we should have
\begin{equation}
J_{n_1}^{-1} \: {}^{t\!}\Gamma^{}_1 \: J_{n_1} = \Gamma^{-1}_1, 
\qquad \qquad
\Gamma^{}_{2s+1-\alpha} = \Gamma^{-1}_\alpha, 
\qquad \alpha = 2,\ldots,2s-1.   
\label{e:4.30}
\end{equation}
To verify these relations we use that
\[
\gamma^{-1} = \psi^{-1}_\infty = I_n + \sum^r_{i=1} 
\sum^{2s-1}_{k=1} h^{2k} \: (h^{-1} \: B^{-1} \: {}^{t\!}P_i 
\: B \: h) \: h^{-2k},
\]
therefore we find the following expression of $\Gamma^{-1}_1$ in 
terms of quasi-periodic quantities,
\[
\Gamma^{-1}_1 = I_{n_1} - \sum^r_{i,j=1} \wt u_{i,1} \: \mu^{2s-1}_j
\: (\wt R^{-1}_1)_{j i} \: {}^{t\!} \wt u_{j,1} \: J_{n_1}.
\]
Comparing now this expression with what we have for $\Gamma_1$ above,
we conclude that the first relation of equations (\ref{e:4.30}) is
satisfied. 

The expressions just given above allow writing a remarkable 
determinant representation for the functions $\Gamma_\alpha$.
It can be shown that
\[
\Gamma_\alpha = \frac{\det(\wt R_\alpha 
+ \wt u_{2s+1-\alpha} \: {}^{t\!} \wt u_\alpha)}{\det \wt R_{\alpha}},
\qquad \alpha = 2,\ldots,s,
\] 
and
\[
\Gamma_\alpha = \frac{\det(\wt R_\alpha 
- \wt u_{2s+1-\alpha} \: {}^{t\!} \wt u_\alpha)}{\det \wt R_{\alpha}},
\qquad \alpha = s+1,\ldots,2s-1.
\] 
Using the properties (\ref{e:4.27}) and (\ref{e:4.28}) we can see
\[
\Gamma_\alpha = \frac{\det \wt R_{\alpha+1}}{\det \wt R_\alpha},
\qquad \alpha = 2,\ldots,s,s+1,\ldots,2s-1.
\]
For these functions we can also easily demonstrate that
\[
\Gamma_{2s+1-\alpha} =  
\frac{\det \wt R_{2s+2-\alpha}}{\det \wt R_{2s+1-\alpha}} =
\frac{\det ({}^{t\!}\wt R_{\alpha})}{\det ({}^{t\!}\wt R_{\alpha+1})} 
= \frac{\det \wt R_{\alpha}}{\det \wt R_{\alpha+1}} = 
\Gamma_\alpha^{-1}, 
\]
using for this purpose the relations (\ref{e:4.28}).
Hence all equations (\ref{e:4.30}) are fulfilled. 

We remember also that for the case $n_1=1$ corresponding to the 
second type of abelian twisted loop Toda equations considered here 
($n = p = 2s-1$), we have $I_{1} = J_{1} = 1$, and so, we can write 
for the function $\Gamma_1$ the expression
\[
\Gamma_1 = 1 - {}^{t\!} \wt u_1 \: M^{2s-1} \: \wt R_1^{-1} \: \wt u_1,
\]  
where $\wt u_1$ is also an $r$-dimensional column vector. It can be
shown that
\[
\Gamma_1 = \frac{\det(\wt R_1 - \wt u_1 \: {}^{t\!} u_1 \: M^{2s-1})}
{\det \wt R_1},
\]
We obtain from the expressions of $\wt R_1$ and $\wt R_2$ directly that  
\[
(\wt R_1)_{i j} \: \mu_j^{-N} = {}^{t\!} \wt u_{i,1} \: J_{n_1} \:
\wt u_{j,1} + \mu_i^{-N} \: (\wt R_2)_{i j}, 
\]
and so, for $n_1 = 1$ we can write
\[
M^{-2s+1} \: \wt R_2 + \wt u_1 \: {}^{t\!} \wt u_1 
= \wt R_1 \: M^{-2s+1}.
\]
Using this relation in the above expression for $\Gamma_1$ as the
ratio of determinants, we easily derive
\begin{equation}
\Gamma_1 = \frac{\det \wt R_2}{\det \wt R_1}.
\label{e:4.31}
\end{equation}
But we have from the equality (\ref{e:4.29}) that
\[
\wt R_1 = - {}^{t\!} \wt R_2,
\]
and so, for $n_1 = 1$ the expression 
(\ref{e:4.31}) gives
\[
\Gamma_1 = (-1)^r.
\]

\section{Soliton solutions} \label{s:5}

\subsection{Odd-dimensional case} \label{s:5.1}

Here we consider the case of $n = p = N = 2s-1$. It means also that
we have $n_1 = 1$. The eigenvectors of the matrices ${}^{t\!}c_-$, 
${}^{t\!}c_+$, $c_-$ and $c_+$ are $n$-dimensional column vectors 
$\Psi_\rho$, $\rho = 1,\ldots,2s-1$, satisfying the relations
\[
{}^{t\!}c_- \Psi_\rho = m \: \epsilon^{s+2\rho}_{2N} \: \Psi_\rho,
\quad
{}^{t\!}c_+ \Psi_\rho = m \: \epsilon^{-s-2\rho}_{2N} \: \Psi_\rho,
\quad
c_- \Psi_\rho = m \: \epsilon^{-s-2\rho}_{2N} \: \Psi_\rho,
\quad
c_+ \Psi_\rho = m \: \epsilon^{s+2\rho}_{2N} \: \Psi_\rho,
\]
where the $2s-1$ components of $\Psi_\rho$ are defined as
\begin{align*}
& (\Psi_\rho)_\alpha = \epsilon^{\alpha(s+2\rho)}_{2N}, &&
\alpha = 1,\ldots,s, \\
& (\Psi_\rho)_\alpha 
= (-1)^{\alpha-s-1} \: \epsilon^{\alpha(s+2\rho)}_{2N}, &&
\alpha = s+1,\ldots,2s-1.
\end{align*}
Consequently, we can give a concrete expression to the formal 
solution (\ref{e:4.26}) as
\begin{gather*}
u_{i,\alpha} = \sum^{2s-1}_{\rho=1} c_{i\rho} \: 
\epsilon^{\alpha(s+2\rho)}_{2N} \: \rme^{-Z_\rho(\mu_i)}, \qquad 
\alpha = 1,\ldots,s \\
u_{i,\alpha} = \sum^{2s-1}_{\rho=1} c_{i\rho} \: (-1)^{\alpha-s-1} 
\: \epsilon^{\alpha(s+2\rho)}_{2N} \: \rme^{-Z_\rho(\mu_i)}, 
\qquad \alpha = s + 1,\ldots,2s-1,
\end{gather*}
where $c_{i\rho}$ are arbitrary constants and we have introduced 
the notation
\[
Z_\rho(\mu_i) = m \: (\epsilon^{-s-2\rho}_{2N} \: \mu_i^{-1} \: z^- 
+ \epsilon^{s+2\rho}_{2N} \: \mu_i^{} \: z^+).
\]
Then, after some calculation using, in particular, properties of 
$\epsilon_{2N}$, we write for the matrix elements of $\wt R_\alpha$ 
for $\alpha \ge 2$: 
\begin{equation}
(\wt R_\alpha)_{i j} = (-1)^\alpha \: \mu_i^{2s+1-\alpha} \: 
\mu_j^{\alpha} \sum_{\rho,\sigma=1}^{2s-1} c_{i\rho} \: c_{j\sigma} \:
\frac{\epsilon^{4\rho + 1 - 2(\rho - \sigma)\alpha}_{2N}}
{1 + \mu^{}_j \: \mu_i^{-1} \epsilon^{-2(\rho - \sigma)}_{2N}} \:
\rme^{- Z_\rho(\mu_i) - Z_\sigma(\mu_j)}.
\label{e:5.1}
\end{equation}
It is clear that to obtain nontrivial solutions to the Toda equations
we should require that at least two coefficients $c_{i\rho}$ for any
$i = 1,\ldots,r$ are different from zero. In this, we construct
solutions depending on only $r$ combinations of independent variables
$z^-$ and $z^+$. We denote such nonzero constants by $C_{J_i}$ and 
$C_{K_i}$. The expression for the matrix elements (\ref{e:5.1}) takes 
then the form
\[
(\wt R_\alpha)_{i j} = (-1)^\alpha \: \mu_i^{2s+1-\alpha} \: 
\epsilon_{2N}^{4J_i + 1 - 2 \alpha J_i} \: C_{J_i} \:
\rme^{-Z_{J_i}(\mu_i)} \: (\wt R'_\alpha)^{}_{i j} \: \mu^\alpha_j
\: C_{J_j} \: \epsilon^{2\alpha J_j}_{2N} \: \rme^{-Z_{J_j}(\mu_j)}, 
\] 
where
\begin{multline*}
(\wt R'_\alpha)_{i j} = \frac{1}
{1 + \mu^{}_j \: \mu^{-1}_i \: \epsilon_{2N}^{2(J_j - J_i)}}
+ \frac{C_{K_j}}{C_{J_j}} \frac{\epsilon_{2N}^{2(K_j - J_j)\alpha}}
{1 + \mu^{}_j \: \mu^{-1}_i \: \epsilon_{2N}^{2(K_j - J_i)}}
\rme^{Z_{J_j}(\mu_j) - Z_{K_j}(\mu_j)} \\
+ \frac{C_{K_i}}{C_{J_i}} 
\frac{\epsilon_{2N}^{4(K_i - J_i) - 2(K_i - J_i)\alpha}}
{1 + \mu^{}_j \: \mu^{-1}_i \: \epsilon_{2N}^{2(J_j - K_i)}}
\rme^{Z_{J_i}(\mu_i) - Z_{K_i}(\mu_i)} \\
+ \frac{C_{K_i} C_{K_j}}{C_{J_i} C_{J_j}} 
\frac{\epsilon_{2N}^{4(K_i - J_i) - 2(K_i - J_i + K_j - J_j)\alpha}}
{1 + \mu^{}_j \: \mu^{-1}_i \: \epsilon_{2N}^{2(K_j - K_i)}}
\rme^{Z_{J_i}(\mu_i) - Z_{K_i}(\mu_i) + Z_{J_j}(\mu_j) - Z_{K_j}(\mu_j)}.
\end{multline*}
It is easy to show that
\[
\Gamma_\alpha = \frac{\det \wt R_{\alpha + 1}}{\det \wt R_\alpha}
= (-1)^r \frac{\det \wt R'_{\alpha + 1}}{\det \wt R'_{\alpha}}.
\]
Recalling also that $\Gamma_1 = (-1)^r$, we see that we can take
$\wt R'_\alpha$ instead of $\wt R_\alpha$ to construct solutions
of the Toda equations using for that the above determinant
representation.

Defining a new set of parameters 
\begin{gather*}
\rho_i = J_i - K_i, \qquad \theta_{\rho_i} = \frac{\pi \rho_i}{2s-1}, 
\qquad \kappa_{\rho_i} 
= - \rmi (\epsilon^{\rho_i}_{2N} - \epsilon^{-\rho_i}_{2N}) 
= 2 \sin \theta_{\rho_i}, \\
\exp{\delta_i} = \frac{C_{K_i}}{C_{J_i}}, \qquad 
\zeta_i = \rmi \: \epsilon^{s + J_i + K_i}_{2N} \: \mu_i, \qquad
f_i = \epsilon_{2N}^{\rho_i} \: \zeta_i, \qquad
\wt f_i = \epsilon_{2N}^{-\rho_i} \: \zeta_i,
\end{gather*}
and introducing the notation
\[
D_{i j}(f,g) = \frac{f_i}{f_i + g_j},
\] 
we can rewrite the expression for $\wt R'_\alpha$ as
\begin{gather}
(\wt R'_\alpha)^{}_{i j} = D_{i j}(f,f) + 
\epsilon_{2N}^{2\rho_i(\alpha-1)} 
\rme^{Z_i(\zeta) + \delta_i - 2\rmi\theta_{\rho_i}} D_{i j}(\wt f, f) 
+ D_{i j}(f, \wt f) \rme^{Z_j(\zeta) + \delta_j - 2\rmi\theta_{\rho_j}}
\epsilon_{2N}^{-2\rho_j(\alpha-1)} \notag \\*
+ \epsilon_{2N}^{2\rho_i(\alpha-1)} 
\rme^{Z_i(\zeta) + \delta_i - 2\rmi\theta_{\rho_i}} 
D_{i j}(\wt f,\wt f) 
\rme^{Z_j(\zeta) + \delta_j - 2\rmi\theta_{\rho_j}}
\epsilon_{2N}^{-2\rho_j(\alpha-1)},
\label{e:5.2}
\end{gather}
where now the dependence on independent variables is given through
\[
Z_i(\zeta) = m \: \kappa_{\rho_i} \: 
(\zeta_i^{-1} \: z^- + \zeta_i^{} \: z^+).
\]
In fact, it appears that it is appropriate to use the matrices 
$T_\alpha = D^{-1}(f,f) \: \wt R'_\alpha$
and write the solutions under construction as
\[
\Gamma_\alpha = \frac{\det T_{\alpha+1}}{\det T_\alpha}.
\]
The problem of constructing multi-soliton solutions for 
the Toda equations (\ref{e:3.7}) is thus reduced 
to calculating the determinant of the $r \times r$ 
matrix $T_\alpha$. 

To obtain a one-soliton solution we set $r=1$. In this case 
$T_\alpha$ are ordinary functions, and we easily find 
\begin{equation}
T_{\alpha + 1} = 
1 + 2\frac{\cos(2\alpha-1)\theta_\rho}{\cos\theta_\rho} 
\rme^{Z(\zeta) + \delta - 2\rmi\theta_\rho}
+ \rme^{2(Z(\zeta) + \delta - 2\rmi\theta_\rho)}.
\label{e:5.3}
\end{equation}
Setting $r = 2$ we work out the determinant of the respective 
$2 \times 2$ matrix explicitly given above and thus obtain for
the two-soliton solution the expression
\begin{eqnarray}
\displaystyle
\det T_{\alpha + 1} &=& 
1 + 2\frac{\cos(2\alpha-1)\theta_{\rho_1}}{\cos \theta_{\rho_1}}
\rme^{\wt Z_1} 
+ 2\frac{\cos (2\alpha-1)\theta_{\rho_2}}{\cos\theta_{\rho_2}}
\rme^{\wt Z_2} + \rme^{2 \wt Z_1} 
+ \rme^{2 \wt Z_2} \notag \\*
&& + \left( 2\eta^{+}_{12} 
\frac{\cos (2\alpha-1)(\theta_{\rho_1} - \theta_{\rho_2})}
{\cos\theta_{\rho_1}\cos\theta_{\rho_2}}
+ 2\eta^{-}_{12} 
\frac{\cos (2\alpha-1)(\theta_{\rho_1} + \theta_{\rho_2})}
{\cos\theta_{\rho_1} \cos\theta_{\rho_2}} 
\right) \rme^{\wt Z_1 + \wt Z_2} \notag \\*
&& + 2 \eta^{+}_{12} \eta^{-}_{12} 
\left(\frac{\cos (2\alpha-1)\theta_{\rho_1}}{\cos \theta_{\rho_1}}
\rme^{\wt Z_1 + 2\wt Z_2} + 
\frac{\cos (2\alpha-1)\theta_{\rho_2}}{\cos \theta_{\rho_2}}
\rme^{2\wt Z_1 + \wt Z_2} \right) \notag \\* 
&& + \left( \eta^{+}_{12} \eta^{-}_{12} \right)^2 
\rme^{2(\wt Z_1 + \wt Z_2)},    
\label{e:5.4} 
\end{eqnarray}
with the `soliton interaction factors' 
\[
\displaystyle 
\eta^{+}_{12} = 
\frac{\displaystyle
({\zeta^{}_1}{\zeta^{-1}_2} 
+ {\zeta^{}_2}{\zeta^{-1}_1}) 
+ 2\cos(\theta_{\rho_1} + \theta_{\rho_2})}
{\displaystyle
({\zeta^{}_1}{\zeta^{-1}_2}
+ {\zeta^{}_2}{\zeta^{-1}_1}) 
+ 2\cos(\theta_{\rho_1} - \theta_{\rho_2})}, 
\quad
\eta^{-}_{12} = 
\frac{\displaystyle
({\zeta^{}_1}{\zeta^{-1}_2}
+ {\zeta^{}_2}{\zeta^{-1}_1}) 
- 2\cos(\theta_{\rho_1} - \theta_{\rho_2})}
{\displaystyle
({\zeta^{}_1}{\zeta^{-1}_2}
+ {\zeta^{}_2}{\zeta^{-1}_1}) 
- 2\cos(\theta_{\rho_1} + \theta_{\rho_2})}, 
\]
and the constant parameters
\[
\rme^{\delta'_1} = 
\frac{\displaystyle
(f_1 + f_2)(\widetilde{f}_1 - f_2)}{(f_1 - f_2)(\widetilde{f}_1 + f_2)},
\qquad
\mathrm{e}^{\delta'_2} = 
\frac{\displaystyle
(f_1 + f_2)(f_1 - \widetilde{f}_2)}{(f_1 - f_2)(f_1 + \widetilde{f}_2)}
\]
giving rise to a shift in the exponents as 
\[
\wt Z_i = Z_i(\zeta) + \delta_i + \delta'_i - 2\rmi\theta_{\rho_i}. 
\]
It can also be shown that performing the corresponding change of 
variables as suggested in the paper \cite{NirRaz08}, one can 
reach the same result along the lines of the Hirota's approach. Here, 
the quantities $\det T_{\alpha + 1}$ constructed by means of the rational 
dressing formalism, will coincide with the Hirota's $\tau$-functions
$\tau_\alpha$, see the paper \cite{NirRaz08} where 
such correspondence was established for the untwisted case.

Now, considering $s = 2$, so that $N = 3$, we describe the 
Dodd--Bullough--Mikhailov equation from Section \ref{s:3.2}. 
Here we have $\Gamma_1 = (-1)^r$, $\Gamma_3^{} = \Gamma_2^{-1}$, 
and so, the mapping $\gamma$ is parameterized by the only nontrivial 
function $\Gamma_2$, denoted here by $\Gamma$. The corresponding soliton 
solutions can easily be derived from the relations (\ref{e:5.3}) and 
(\ref{e:5.4}) putting $\alpha = 2$ and $\alpha = 3$ in order and
taking into account that $\theta_{\rho} = \pi\rho/3$. Remember here 
that $\rho = J - K$, where $J$ and $K$ take values $1$, $2$ or $3$ 
only. In particular, it is easy to see that the one-soliton solution 
can be written as
\[
\Gamma = \frac{1 
- 4 \: \rme^{Z(\zeta) + \delta - 2 \rmi\theta_\rho} 
+ \rme^{2(Z(\zeta) + \delta - 2 \rmi\theta_\rho)}}
{(1 + \: \rme^{Z(\zeta) + \delta - 2 \rmi\theta_\rho})^2}.
\] 
For the two-soliton solution, we should respectively simplify
the expression (\ref{e:5.4}). The corresponding expressions reproduce 
the one- and two-soliton solutions of the Dodd--Bullough--Mikhailov 
equation obtained in the paper \cite{AssFer07} by means of the 
Hirota's method.

\subsection{Even-dimensional case} \label{s:5.2}

Here we consider the case of $n = 2s$, with  $p = N = 2s-1$. It means 
also that now we have $n_1 = 2$. The eigenvectors of the matrices 
${}^{t\!}c_-$, ${}^{t\!}c_+$, $c_-$ and $c_+$ are $2s$-dimensional 
column vectors $\Psi_\rho$, $\rho = 1,\ldots,2s-1$, satisfying the 
relations
\[
{}^{t\!}c_- \Psi_\rho = m \: \epsilon^{s+2\rho}_{2N} \: \Psi_\rho,
\quad
{}^{t\!}c_+ \Psi_\rho = m \: \epsilon^{-s-2\rho}_{2N} \: \Psi_\rho,
\quad
c_- \Psi_\rho = m \: \epsilon^{-s-2\rho}_{2N} \: \Psi_\rho,
\quad
c_+ \Psi_\rho = m \: \epsilon^{s+2\rho}_{2N} \: \Psi_\rho,
\]
where we define the $2s$ components of $\Psi_\rho$ as
\[
(\Psi_\rho)_0 = (\Psi_\rho)_1 = \epsilon^{\alpha(s+2\rho)}_{2N},
\qquad
(\Psi_\rho)_\alpha = \sqrt{2} \: \epsilon^{\alpha(s+2\rho)}_{2N}, 
\qquad \alpha = 2,\ldots,s
\]
and
\[
(\Psi_\rho)_\alpha 
= (-1)^{\alpha-s-1} \: \sqrt{2} \: \epsilon^{\alpha(s+2\rho)}_{2N}, 
\qquad \alpha = s+1,\ldots,2s-1.
\]
Besides, respective to the only zero eigenvalue, $c_-$, $c_+$ and 
their transposed matrices have one and the same null-vector that 
can be defined as ${}^{t\!}\Psi_0 = (1,-1,0,\ldots,0)$. 

Consequently, the solution (\ref{e:4.26}) takes the form
\[
(u_{i,1})_0 = c_{i0} + \sum_{\rho=1}^{2s-1} c_{i\rho} \:
\epsilon_{2N}^{s + 2\rho} \: \rme^{-Z_{\rho}(\mu_i)}, \qquad
(u_{i,1})_1 = - c_{i0} + \sum_{\rho=1}^{2s-1} c_{i\rho} \:
\epsilon_{2N}^{s + 2\rho} \: \rme^{-Z_{\rho}(\mu_i)}, 
\]
and
\begin{gather*}
u_{i,\alpha} = \sum^{2s-1}_{\rho=1} c_{i\rho} \: \sqrt{2} \: 
\epsilon^{\alpha(s+2\rho)}_{2N} \: \rme^{-Z_\rho(\mu_i)}, \qquad 
\alpha = 2,\ldots,s \\
u_{i,\alpha} = \sum^{2s-1}_{\rho=1} c_{i\rho} \: (-1)^{\alpha-s-1} 
\: \sqrt{2} \: \epsilon^{\alpha(s+2\rho)}_{2N} \: \rme^{-Z_\rho(\mu_i)}, 
\qquad \alpha = s + 1,\ldots,2s-1,
\end{gather*}
where $c_{i0}$ and $c_{i\rho}$ are arbitrary constants and, as usual,
we have introduced the notation
\[
Z_\rho(\mu_i) = m \: (\epsilon^{-s-2\rho}_{2N} \: \mu_i^{-1} \: z^- 
+ \epsilon^{s+2\rho}_{2N} \: \mu_i^{} \: z^+).
\]
Note that $u_{i,1}$ is now a $2$-dimensional column vector with
the components $(u_{i,1})_0$ and $(u_{i,1})_1$ given in order.

For the quasi-periodic quantities $\wt R_\alpha$ introduced in
Section \ref{s:4} we obtain the expressions
\begin{align*}
(\wt R_1)_{i j} = &- \frac{2 \: \mu_i^{2s} \: \mu_j^{2s}}
{\mu_i^{2s-1} + \mu_j^{2s-1}} \: c_{i0} \: c_{j0}  \\* 
& + 2 \: \mu_i^{} \: \mu_j^{2s} 
\sum_{\rho,\sigma=1}^{2s-1} c_{i\rho} \: c_{j\sigma} \:
\frac{\epsilon^{2(s + \rho + \sigma)}_{2N}}
{1 + \mu^{}_j \: \mu_i^{-1} \epsilon^{-2(\rho - \sigma)}_{2N}} \:
\rme^{- Z_\rho(\mu_i) - Z_\sigma(\mu_j)},  \\*
(\wt R_\alpha)_{i j} = & \frac{2 \: \mu_i^{2s} \: \mu_j^{2s}}
{\mu_i^{2s-1} + \mu_j^{2s-1}} \: c_{i0} \: c_{j0} \\*
 & + 2 \: (-1)^\alpha \: \mu_i^{2s+1-\alpha} \: \mu_j^{\alpha} 
\sum_{\rho,\sigma=1}^{2s-1} c_{i\rho} \: c_{j\sigma} \:
\frac{\epsilon^{4\rho + 1 - 2(\rho - \sigma)\alpha}_{2N}}
{1 + \mu^{}_j \: \mu_i^{-1} \epsilon^{-2(\rho - \sigma)}_{2N}} \:
\rme^{- Z_\rho(\mu_i) - Z_\sigma(\mu_j)},
\end{align*}
that are to be used in the determinant representation derived
earlier for constructing the soliton solutions.

It is easy to see that if we set here $c_{i0} = 0$, then we 
come to the same solutions given for $n = 2s-1$ by the relations
(\ref{e:5.2})--(\ref{e:5.4}), with the only unessential 
difference that for $n = 2s$ the rational dressing gives 
$\Gamma_1 = (-1)^r J^r_2$. Therefore, to obtain new solutions 
we consider that in what follows $c_{i0}$ does not vanish.
 
To construct such new simplest soliton solutions we thus
assume that for each value of the index $i$ only one arbitrary
constant $c_{i\rho}$, apart from the $c_{i0}$, is different from
zero. To keep up with the notations used in the preceding section,
we denote such nonvanishing coefficients by $C_{I_i}$ and $C_{0_i}$.
Then we can write for the above $r \times r$ matrices $\wt R_1$
and $\wt R_\alpha$
\[
(\wt R_1)_{i j} = - 2 \: \mu^{}_i \: C_{0_i} \: 
(\wt R'_1)_{i j} \: C_{0_j} \: \mu^{2s}_j, \qquad
(\wt R_\alpha)_{i j} = 2 \: \mu^{}_i \: C_{0_i} \: 
(\wt R'_\alpha)_{i j} \: C_{0_j} \: \mu^{2s}_j,
\]
where $\wt R'_1$ and $\wt R'_\alpha$ can be represented as
\begin{gather*}
(\wt R'_1)_{i j} = D_{i j}(\zeta^{2s-1},\zeta^{2s-1})
- \rme^{- Z'_i} \: D_{i j}(\zeta,\zeta) \: \rme^{- Z'_j}, \\
(\wt R'_\alpha)_{i j} = D_{i j}(\zeta^{2s-1},\zeta^{2s-1})
- (-1)^\alpha \: \zeta^{2s - \alpha}_i \: 
\rme^{- Z'_i} \: D_{i j}(\zeta,\zeta) \: \rme^{- Z'_j}
\: \zeta^{\alpha - 2s}_j.
\end{gather*}
Here we use the same notation for the matrices $D(f,g)$ introduced
in the preceding Section \ref{s:5.1}, and besides,
\[
Z'_i = Z_{i}(\zeta) - \delta_i - \rmi\theta_{s+2I_i}, \qquad
Z_{i}(\zeta) = m \: (\zeta^{-1}_i \: z^- + \zeta^{}_i \: z^+),
\]
with the set of parameters
\[ 
\zeta_i = \epsilon^{s + 2I_i}_{2N} \: \mu_i, \qquad
\rme^{\delta_i} = \frac{C_{I_i}}{C_{0_i}}, \qquad
\theta_{s+2I_i} = \frac{\pi (s + 2I_{i})}{2s - 1}.
\]
We also rewrite the explicit forms of the components of the
$2$-dimensional column vector $\wt u_{i,1}$ in terms of the
notations introduced above. We have
\[
(\wt u_{i,1})_0 = \mu_i \: C_{0_i} \: (1 + \exp(- Z'_i)),
\qquad
(\wt u_{i,1})_1 = - \mu_i \: C_{0_i} \: (1 - \exp(- Z'_i)).
\]
Hence, according to the general relations derived in Section 
\ref{s:4}, we can take the matrices  
$T_\alpha = D^{-1}(\zeta^{2s-1},\zeta^{2s-1}) \wt R'_\alpha$ 
instead of $\wt R_\alpha$ and write for the solutions of the 
Toda equations (\ref{e:3.12}) the following expressions:
\[
\Gamma_1 = I_2 + \sum_{i,j=1}^r v_i \: 
(\wt R^{\prime -1}_1)_{i j} 
\: {}^{t\!} v_j \: J_2,
\]
where $v_i$ are $2$-dimensional column vectors with the components
\[
v_{i,0} = \frac{1}{\sqrt{2}} (1 + \exp{(- Z'_i)}), \qquad
v_{i,1} = - \frac{1}{\sqrt{2}} (1 - \exp{(- Z'_i)}),
\]
and
\[
\Gamma_\alpha = \frac{\det T_{\alpha+1}}{\det T_{\alpha}},
\qquad \alpha = 2,\ldots,s.
\]

To obtain a one-soliton solution of the type under consideration, 
we put $r = 1$, for which $T_\alpha$ are ordinary functions. It 
is easy to show that in this case we have
\[
\Gamma_1 = \left( \begin{array}{cc}
0 & \Gamma \\ 
\Gamma^{-1} & 0
\end{array} \right), \qquad 
\Gamma = \frac{1 + \exp(- Z')}{1 - \exp(- Z')},
\]
and
\[
\Gamma_\alpha = \frac{1 + (-1)^\alpha \: \exp(- 2 Z')}
{1 - (-1)^\alpha \: \exp(- 2 Z')}, \qquad 
\alpha = 2,\ldots,s.
\]
Note that apart from the relation 
$\Gamma_{2s+1-\alpha}^\ph = \Gamma_\alpha^{-1}$ here we 
also have $\Gamma_{\alpha+1}^\ph = \Gamma_\alpha^{-1}$.
It is clear that to have a mapping $\gamma$ belonging to
$G_0$ we should take $\Gamma_1 \: J_2$ instead of the
above $\Gamma_1$.

Setting $r = 2$ we work out the corresponding $2 \times 2$ 
matrices and thus obtain new two-soliton solutions to 
(\ref{e:3.12}). The calculations lead to the expressions
\[
\Gamma_1 = \left( \begin{array}{cc}
\Gamma & 0 \\ 
0 & \Gamma^{-1}
\end{array} \right), \qquad 
\Gamma = \frac{1 + \rme^{- \wt Z_1} - \rme^{- \wt Z_2} 
- \eta_{12} \: \rme^{-(\wt Z_1 + \wt Z_2)}}
{1 - \rme^{- \wt Z_1} + \rme^{- \wt Z_2} 
- \eta_{12} \: \rme^{-(\wt Z_1 + \wt Z_2)}},
\]
where the `soliton interaction factor' is now 
\[
\eta_{12} = \frac{\zeta_1 - \zeta_2}{\zeta_1 + \zeta_2}
\cdot \frac{\zeta^{2s-1}_1 - \zeta^{2s-1}_2}
{\zeta^{2s-1}_1 + \zeta^{2s-1}_2},
\]
and we have introduced a new parameter $\delta'$ defined
by 
\[
\rme^{\delta'} = \frac{\zeta^{2s-1}_1 + \zeta^{2s-1}_2}
{\zeta^{2s-1}_1 - \zeta^{2s-1}_2}
\] 
and producing a shift in the exponents,
\[
\wt Z_i = Z'_i - \delta' 
= Z_i(\zeta) - \delta_i - \delta' - \rmi \theta_{s+2I_i}.
\]
We also have
\begin{multline*}
\det T_{\alpha + 1} = 1 + (-1)^\alpha \: (\rme^{-2 \wt Z_1} 
+ \rme^{-2 \wt Z_2}) \\ 
- 4 \: (-1)^\alpha \: 
\frac{\zeta_1^{\alpha} \zeta_2^{2s-\alpha} 
+ \zeta_2^{\alpha} \zeta_1^{2s-\alpha}}
{(\zeta_1^{} + \zeta_2^{})(\zeta_1^{2s-1} + \zeta_2^{2s-1})} \:
\rme^{-(\wt Z_1 + \wt Z_2)} 
+ \eta_{12}^2 \: \rme^{-2 (\wt Z_1 + \wt Z_2)}.
\end{multline*}
Note finally that under the permutation of the parameters $\zeta_1$
and $\zeta_2$ the function $\Gamma$ transforms into $\Gamma^{-1}$, 
thus $\Gamma_1$ goes to $\Gamma_1^{-1}$, while $\Gamma_\alpha$ for 
the other values of $\alpha$ all stay invariant.

\section{Conclusion}

We have considered the abelian Toda systems associated with the 
loop groups of the complex general linear groups. Using the method 
of rational dressing, along the lines of \cite{NirRaz08}, we have 
constructed soliton solutions to these equations in the twisted cases, 
that is, when the gradations are generated by outer automorphisms of 
the structure Lie algebras. Our consideration can be generalized to 
Toda systems connected with other loop groups, such as twisted and 
untwisted loop groups of the complex orthogonal and symplectic groups.
It is worth noting that, as we have already observed here, the pole 
positions of the dressing meromorphic mappings and their inverse ones
turn out to be bound up with each other because of the specific 
structure of the outer automorphism leading to the twisted cases. 
This circumstance made part of the formulae more intricate than in 
the untwisted general linear case considered in the preceding paper
\cite{NirRaz08}. Actually, similar problems of coinciding pole
positions arise also due to the specific group conditions. 
We will address to this problem and present our respective 
results in some future publications.

\vskip2mm
This work was supported in part by the Russian Foundation for Basic
Research under grant \#07--01--00234 and by the joint DFG--RFBR grant
\#08--01--91953. One of the authors (A.V.R.) wishes to acknowledge
the warm hospitality of the Erwin Schr\"o\-dinger International
Institute for Mathematical Physics where a part of this work was carried out.

\end{document}